\documentclass[12pt,preprint]{aastex}

\newcommand{\dmm}{\mbox{$\Delta m_{15}(B)$}}
\newcommand{\kms}{\mbox{km s$^{-1}$}}

\shorttitle{SN 2001ay}
\shortauthors{Krisciunas et al.}

\begin{document}

\title{The Most Slowly Declining Type Ia Supernova 2001ay}

\author{Kevin Krisciunas,\altaffilmark{1}
Weidong Li,\altaffilmark{2}
Thomas Matheson,\altaffilmark{3}
D. Andrew Howell,\altaffilmark{4,5}
Maximilian Stritzinger,\altaffilmark{6,7}
Greg Aldering,\altaffilmark{8}
Perry L. Berlind,\altaffilmark{9}
M. Calkins,\altaffilmark{9}
Peter Challis,\altaffilmark{10}
Ryan Chornock,\altaffilmark{10}
Alexander Conley,\altaffilmark{111}
Alexei V. Filippenko,\altaffilmark{2}
Mohan Ganeshalingam,\altaffilmark{2}
Lisa Germany,\altaffilmark{12}
Sergio Gonz\'{a}lez,\altaffilmark{13}
Samuel D. Gooding,\altaffilmark{1}
Eric Hsiao,\altaffilmark{8}
Daniel Kasen,\altaffilmark{14}
Robert P. Kirshner,\altaffilmark{10}
G. H. ``Howie'' Marion,\altaffilmark{10}
Cesar Muena,\altaffilmark{13}
Peter E. Nugent,\altaffilmark{8}
M. Phelps,\altaffilmark{9}
Mark M. Phillips,\altaffilmark{13}
Yulei Qiu,\altaffilmark{15}
Robert Quimby,\altaffilmark{16}
K. Rines,\altaffilmark{9}
Jeffrey M. Silverman,\altaffilmark{2}
Nicholas B. Suntzeff,\altaffilmark{1}
Rollin C. Thomas,\altaffilmark{8}
and Lifan Wang\altaffilmark{1}
}


\altaffiltext{1}{Department of Physics \& Astronomy, George P. and
  Cynthia Woods Mitchell Institute for Fundamental Physics \&
  Astronomy, Texas A\&M University, 4242 TAMU, College Station, TX
  77843-4242; {krisciunas@physics.tamu.edu},
  {suntzeff@physics.tamu.edu}, {sam.gooding86@gmail.com},
  {wang@physics.tamu.edu}. }

\altaffiltext{2}{Department of Astronomy, University of California,
  Berkeley, CA 94720-3411; {weidong@astro.berkeley.edu},
  {alex@astro.berkeley.edu}, {mganesh@astro.berkeley.edu},
  {jsilverman@astro.berkeley.edu}. }

\altaffiltext{3}{National Optical Astronomy Observatories, 950
  N. Cherry Avenue, Tucson, AZ 85719-4933; {matheson@noao.edu}. }

\altaffiltext{4}{Las Cumbres Observatory Global Telescope Network,
  6740 Cortona Drive, Suite 102, Goleta, CA 93117;
  {ahowell@lcogt.net}. }

\altaffiltext{5}{Department of Physics, University of California, 
Santa Barbara, CA 93106-9530. }

\altaffiltext{6}{The Oskar Klein Centre, Department of Astronomy,
  Stockholm University, AlbaNova, 10691 Stockholm, Sweden;
  {max.stritzinger@astro.su.se}. }

\altaffiltext{7}{Dark Cosmology Centre, Niels Bohr Institute,
  University of Copenhagen, Juliane Maries Vej 30, 2100 Copenhagen \O,
  Denmark; {max@dark-cosmology.dk}. }

\altaffiltext{8}{Lawrence Berkeley National Laboratory, 1 Cyclotron Road,
  Berkeley, CA 94720; {galdering@lbl.gov}, {ehsiao@lbl.gov}, 
  {penugent@lbl.gov}, {rcthomas@lbl.gov}. }

\altaffiltext{9}{Fred L. Whipple Observatory, P. O. Box 97, Amado, AZ
  85645; {berlind@cfa.harvard.edu}. }

\altaffiltext{10}{Harvard-Smithsonian Center for Astrophysics, 60
  Garden Street, Cambridge, MA 02138; {pchallis@cfa.harvard.edu},
  {rchornock@cfa.harvard.edu}, {kirshner@cfa.harvard.edu},
  {hman@astro.as.utexas.edu}. }

\altaffiltext{11}{Department of Astronomy, University of Colorado,
  Boulder, CO 80309; {alexander.conley@colorado.edu}. }

\altaffiltext{12}{Swinburne University of Technology, 1 John Street,
  Hawthorn, VIC, 3122, Australia; {lgermany@swin.edu.au}. }

\altaffiltext{13}{Las Campanas Observatory, Casilla 601, La Serena,
  Chile; {mmp@lco.cl}. }

\altaffiltext{14}{Department of Physics, University of California,
  Berkeley, CA 94720; {kasen@berkeley.edu}. }

\altaffiltext{15}{National Astronomical Observatories of China,
  Chinese Academy of Sciences, Beijing 100012, China;
  {qiuyl@bao.ac.cn}. }

\altaffiltext{16}{Department of Astronomy, California Institute of
  Technology, Pasadena, CA 91125; {quimby@astro.caltech.edu}. }

\newpage

\begin{abstract}


We present optical and near-infrared photometry,
as well as ground-based optical spectra and {\it Hubble Space
  Telescope} ultraviolet spectra, of the Type Ia supernova (SN)
2001ay.  At maximum light the Si~II and Mg~II lines indicated
expansion velocities of 14,000 \kms, while Si~III and S~II showed
velocities of 9,000 \kms.  There is also evidence for some unburned
carbon at 12,000 \kms. SN~2001ay exhibited a decline-rate parameter
\dmm\ = 0.68 $\pm$ 0.05 mag; this and the $B$-band photometry at $t
\gtrsim +25$~d past maximum make it the most slowly declining Type Ia SN
yet discovered.  Three of four super-Chandrasekhar-mass candidates
have decline rates almost as slow as this.  After correction for
Galactic and host-galaxy extinction, SN~2001ay had $M_B = -19.19$ and
$M_V = -19.17$ mag at maximum light; thus, it was {\em not}
overluminous in optical bands. In near-infrared bands it was
overluminous only at the 2$\sigma$ level at most.  For a rise
time of 18~d (explosion to bolometric maximum) the
implied $^{56}$Ni yield was (0.58 $\pm$ 0.15)/$\alpha$ M$_{\odot}$,
with $\alpha = L_{\rm max}/E_{\rm Ni}$ probably in the range 1.0 to 1.2.  The 
$^{56}$Ni yield is comparable to that of many Type Ia supernovae.  The 
``normal'' $^{56}$Ni yield and the typical peak optical brightness suggest 
that the very broad optical light curve is explained by the trapping of the
$\gamma$ rays in the inner regions.

\end{abstract}

\section{Introduction}

\citet{Phi93} first established that Type Ia supernovae (SNe) are 
{\em standardizable} candles at optical wavelengths: there is a
correlation between their absolute magnitudes at maximum light and the
rate at which these objects fade. This fact allowed Type Ia SNe to be
used to determine that the expansion of the Universe is currently
accelerating \citep{Rie_etal98, Per_etal99}.  More recently, we have
discovered that in near-infrared (IR) photometric bands Type Ia SNe
are {\em standard} candles \citep{Mei00, Kri_etal04a, WV_etal08,
  Fol_etal10, Man_etal11}.  Except for a fraction of the rapidly
declining Type Ia SNe whose prototype is SN~1991bg \citep{Fil_etal92,
  Lei_etal93}, which peak in the near-IR after the time of $B$-band
maximum, the near-IR absolute magnitudes at maximum light are at most
only slightly dependent on the decline-rate parameter
\dmm\ \citep{Kri_etal09, Fol_etal10}.


We do not understand well how to model a Type Ia SN.  For much of the
previous decade we thought that (most) Type Ia SNe were carbon-oxygen
white dwarfs that approach the Chandrasekhar limit (1.4 M$_{\odot}$)
owing to mass transfer from a nearby companion \citep[for a review,
  see, e.g.,][]{Liv00}.  Some Type Ia SNe might be mergers of two
white dwarfs \citep{Iben_Tut84}.  Over the past few years the
double-degenerate scenario has gained prominence among many
researchers.  Unfortunately, it is not yet possible to determine with
a high degree of probability that a given Type Ia SN is a single or
double-degenerate event, except perhaps for the few that produced more
than a Chandrasekhar mass of fusion products; these would result from
double-degenerate mergers.  An excellent summary of evidence for
different kinds of progenitors of Type Ia SNe is given by
\citet{How11}.

The spectroscopic classification scheme stipulates that Type I SNe
do not exhibit obvious hydrogen in their spectra, and that
Type II SNe do show prominent hydrogen \citep{Min41}; see
\citet{Fil97} for a review.  Type Ia SNe show Si~II in absorption,
blueshifted from its rest wavelength at 6355~\AA.  Two objects have
been shown to have strong Si~II in absorption and H$\alpha$ in
emission: SN~2002ic \citep{Ham_etal03} and SN~2005gj
\citep{Ald_etal06, Pri_etal06a}.  The interpretation is that these
objects are Type Ia SNe interacting with circumstellar material (CSM)
or (less likely) with the general interstellar medium (ISM).

SN~2006X was found to exhibit variable Na~I~D absorption
\citep{Pat_etal07}.  Two other examples showing variable
Na~I absorption are SN~1999cl \citep{Blo_etal09} and SN~2006dd
\citep{Str_etal10}.  Light from the SNe is ionizing the circumstellar
medium.  This leads to recombination and line variations.
It should be pointed out that SNe~1999cl and
2006X have large reddening that may involve multiple scattering as
well as normal dust extinction \citep{Wan05, Goo08}.

It is possible to relate the peak bolometric luminosity of a Type Ia
SN and the implied amount of radioactive $^{56}$Ni produced in the
explosion.  Most events generate between 0.4 and 0.7 M$_{\odot}$ of
$^{56}$Ni \citep{Str_etal06}.  The remarkable SN~2003fg \citep[also
  known as SNLS-03D3bb;][]{How_etal06} was sufficiently overluminous
that it was regarded to have been caused by the explosion of more than
1.4 M$_{\odot}$ of carbon and oxygen.  Three additional
super-Chandrasekhar-mass candidates have recently been found:
SN~2006gz \citep{Hic_etal07}, SN~2007if \citep{Yua_etal10,
  Sca_etal10}, and SN~2009dc \citep{Yam_etal09, Tan_etal10,
  Sil_etal10, Tau_etal11}. As summarized by \citet[][Table
  4]{Sca_etal10}, these objects can be up to a magnitude more luminous
at optical wavelengths than typical Type Ia SNe.  Furthermore, three
of these four have extremely slowly declining light curves
(\dmm\ $\approx 0.7$ mag).  In a double-degenerate system, the more
massive white dwarf can tidally disrupt the less massive white dwarf,
which has a larger radius. A disk or shell of material is created,
retarding the expansion of the more massive white dwarf once it
explodes.  Another object with a very slow decline rate was SN~2005eq,
a target of the Carnegie Supernova Project \citep{Fol_etal10,
  Con_etal10}.

In this paper we present optical and near-IR photometry, plus a
plethora of spectra, of the equally remarkable SN~2001ay.  Preliminary
analysis of the light curve and spectra of SN~2001ay is given by
\citet{Phi_etal03}.  Two {\it Hubble Space Telescope (HST)} STIS spectra
(from 2001 May 2 and May 9) have already been published by
\citet{Fol_etal08}, in a study of possible luminosity indicators
in the ultraviolet (UV) spectra of Type Ia SNe.


\section{Optical and IR Photometry}  

SN~2001ay was discovered on 2001 April 18.4 (UT dates are used
throughout this paper) by \citet{Swi_Li01}. Its position was $\alpha =
14^{\rm h}26^{\rm m}17.0^{\rm s}$, $\delta = +26^\circ 14' 55.8''$
(J2000), some $10.3''$ west and $9.3''$ north of the nucleus of the
spiral galaxy IC 4423.  Basic information on SN~2001ay is given in
Table \ref{properties}.  Figure \ref{finder} shows the host galaxy and
SN, and identifies a number of field stars of interest.

A sizable fraction of our optical photometry of SN~2001ay was
obtained with the 0.76-m Katzman Automatic Imaging Telescope (KAIT) at
Lick Observatory \citep{Li_etal00, Fil_etal01}.  KAIT images are $6.7'
\times 6.7'$ in size, with a scale of $0.80''$ per pixel.  For the
first three nights of photometry with the Lick Observatory Nickel 1-m
telescope, the CCD camera had a chip with $0.30''$ pixels in $2 \times
2$ readout mode and gave a field size of $5.1' \times 5.1'$.
Thereafter, a new chip was installed, having $0.37''$ pixels in $2
\times 2$ readout mode and a field size of $6.3' \times 6.3'$.  The
two different CCD chips employed with the Nickel telescope had
considerably different quantum efficiencies at blue wavelengths.

The $UBVRI$ photometry of some field stars near SN~2001ay is found in
Table \ref{ubvri_field}. Calibration of optical photometry was
accomplished from observations carried out on five photometric nights
(two with the CTIO 0.9-m, two with the Nickel, and one with the CTIO
1.5-m telescopes) using the standard stars of \citet{Lan92}.  A subset
of our photometry (the KAIT data) has already been published by
\citet[][Table 6]{Gan_etal10}, but without K- and S-corrections.  Our
data presented here were reduced in the {\sc iraf}
environment.\footnote[17]{{\sc iraf} is distributed by the National
  Optical Astronomy Observatory, which is operated by the Association
  of Universities for Research in Astronomy, Inc., under cooperative
  agreement with the National Science Foundation (NSF).}
Transformation of the data to the system of \citet{Lan92} was
accomplished using equations equivalent to those of
\citet{Wan_etal09a}.  In their Table 1, the reader will find
characteristic color terms used for the transformations.  Our own
direct determinations of these color terms are entirely consistent
with their values.

Table \ref{jh_field} gives the $J_s$ and $H$-band photometry of some
of the field stars.  We give the averages obtained on seven
photometric nights using the Las Campanas 1-m Swope telescope.  The
calibration was accomplished using a stand-alone version of {\sc
  daophot}, some {\sc fortran} programs written by one of us (N.B.S.),
and the IR standards of \citet{Per_etal98}.  Thus, our near-IR
photometry of SN~2001ay is on the \citet{Per_etal98} photometric
system.

We can compare the photometry of the principal IR secondary standard
(``Star 1'') with data from the Two-Micron All Sky Survey 
\citep[2MASS;][]{Skr_etal06}.
Whereas we obtained $J_s$ = 13.718 $\pm$ 0.014 and $H$ = 13.251 $\pm$
0.011 mag, 2MASS found $J$ = 13.718 $\pm$ 0.027 and $H$ = 13.214 $\pm$
0.030 mag.  These values are statistically in agreement. For ``Star
6'' we obtained $J_s$ = 15.153 $\pm$ 0.021 and $H$ = 14.881 $\pm$
0.008 mag, whereas 2MASS found $J$ = 15.132 $\pm$ 0.045 and $H$ =
14.807 $\pm$ 0.060 mag.

Tables \ref{ubvri} and \ref{jh} give the optical and IR photometry of
SN~2001ay, respectively. Table \ref{ubvri} lists point-spread function
(PSF) magnitudes from KAIT, the first three Nickel 1-m nights, and one
night of the CTIO 1.5-m telescope; the subtracted template images were
obtained with KAIT on 2002 July 9 and 10.  For the single night of
imaging with the CTIO 1.5-m telescope, the $U$-band magnitude was
determined from aperture photometry; no $U$-band template was
available. For the final night of Nickel 1-m photometry, PSF
magnitudes were derived, but without template subtraction.  Aperture
photometry without template subtraction was conducted for all other
optical images.\footnote[18]{Due to the long delay in assembling the
  data and writing this paper, some images are no longer retrievable
  from magnetic storage media. Only aperture photometry was possible
  in 2001, when some of the data were reduced, owing to the lack (at
  the time) of host-galaxy templates.}

To compare the photometry of different SNe, one should transform the
data to the rest-frame equivalent by subtracting the so-called
K-corrections \citep {Oke_San68, Ham_etal93, Kim_etal96, Hog_etal02,
  Nug_etal02}.  For each SN one computes the time since maximum
brightness, and then obtains a rest-frame timescale by dividing the
differential times by $1 + z$, where $z$ is the object's heliocentric
redshift.  We have worked out the optical K-corrections using actual
spectra of SN~2001ay, adopting the \citet{Bes90} filter profiles as
reference; these are given in Table \ref{opt_kcorrs} and shown in
Figure \ref{kcorr}.

To account for differences in SN photometry that result from the use
of different telescopes, CCD chips, and filters, we use the method of
spectroscopic corrections (S-corrections), as outlined by
\citet{Str_etal02} and \citet{Kri_etal03}.  For optical bands we adopt
as reference the filter specifications of \citet{Bes90}. 
{\em At minimum} one constructs the effective filter profiles using 
the laboratory transmission curves of the filters multiplied by an
appropriate atmospheric function that includes atmospheric absorption
lines; then, one multiplies that result by the quantum efficiency of
the CCD chip as a function of wavelength.  Telescope and instrument
optics also contribute to the effective filter profiles.  For the KAIT
and Nickel 1-m corrections we have used an atmospheric
function\footnote[19]{The atmospheric functions combine the mean
broad-band extinction values with atmospheric absorption lines.}
appropriate to Lick Observatory (elevation 1290 m).  For the CTIO
0.9-m data we have used a different atmospheric function appropriate
to Cerro Tololo (elevation 2215 m).  In practice, one arbitrarily
shifts the effective filter profiles toward longer or shorter
wavelengths so that synthetic photometry based on spectra of standard
stars gives color terms that match those obtained from actual
photometry of photometric standards.  We used spectra of 50 stars from
the sample of \citet{Str_etal05} to calculate our synthetic magnitudes
using an {\sc iraf} script written by one of us (N.B.S.).  This script
was then used to calculate the S-corrections based on spectra of
SN~2001ay itself.

In Figure \ref{scorr_bv} we show the $B$- and $V$-band S-corrections
for SN~2001ay.  One {\em adds} these corrections to the photometry.
As one can see in Figure \ref{scorr_bv}, if an object like SN~2001ay
were observed with the CTIO 0.9-m and Nickel 1-m telescopes, there
could be photometric discrepancies as large as $\sim 0.06$ mag in the
$B$ band, depending on the epoch.  In the $V$ band the corresponding
differences are much smaller.

Even after the application of the S-corrections, we find that the
Nickel 1-m $B$-band data based on aperture magnitudes are
systematically 0.09 mag fainter than the KAIT measurements.  
For the first three nights of Nickel imaging we were able to
eliminate these systematic differences by means of template subtraction.
The images subsequently obtained with the Nickel 1-m telescope and its 
newer chip did not lend themselves to image subtraction using the KAIT
templates. (Depending on the number of field stars, the angular size of a
nearby galaxy in the field, and the seeing, the image rescaling and remapping algorithm
can fail.)  As a result, we have devised a third set of corrections to the
photometry based on images with the new chip in the Nickel camera.
Using {\sc addstar} within the {\sc daophot} package of
{\sc iraf}, we were able to add an artificial star to the KAIT 
image-subtraction templates at the pixel location of the SN.  This
artificial star can be scaled to give a standard magnitude of a
desired value such as $B$ = 17.97 and $V$ = 17.14 mag, just about the
brightness of the SN on 2001 May 14 (JD 2,452,043.8), in the middle of
the run of Nickel 1-m aperture photometry.  Artificial stars having the
identical PSF magnitude as the fake star at the location of the SN are
placed at blank places in the KAIT templates using {\sc addstar}. 
Since the Nickel 1-m aperture photometry was typically derived 
with a software aperture of radius 10 pixels (px) and a sky
annulus ranging from 12 to 20 px, we can then compare the aperture
magnitudes of the fake SN with aperture photometry of the fake stars
in the blank places of the image.  We use a software aperture radius
and sky annulus of identical size in {\em arc seconds} to that used
for the Nickel 1-m photometry.  In this way, we can obtain an 
estimate of the systematic errors of the SN aperture photometry.

We determined that the $B$-band magnitudes from Nickel 1-m aperture
photometry of the SN were 0.08 mag fainter than what we would have
measured without the presence of the host galaxy. Similarly, in the
$V$ band we found that the Nickel 1-m aperture magnitudes were fainter
by 0.05 mag. For the two nights of CTIO 0.9-m aperture photometry,
when the SN was more than 1 mag brighter, the SN was too faint by 0.03
mag in $B$ and $V$ without this correction.  Thus, by applying the
S-corrections and these additional corrections, we can reconcile
almost all systematic differences in the $B$- and $V$-band photometry
obtained with different telescopes, using different cameras, and
using different data-reduction methods (i.e., PSF magnitudes with
image-subtraction templates vs. aperture magnitudes without
subtractions).

Further justification of this method comes from a consideration of the
optical photometry from the first Nickel 1-m chip and the photometry
from one night with the CTIO 1.5-m telescope.  Data from these four
nights {\em can} be derived using the KAIT host-galaxy templates.  In
the $B$ band, photometry of the SN using PSF magnitudes and image
subtraction was 0.07 to 0.16 mag brighter than aperture photometry
using an annulus for the subtraction of the sky level.
In the $V$ band, the photometric values of the SN on the
first three Nickel 1-m nights were 0.06 mag brighter than values
derived from aperture photometry.
This result is contrary to typical experience;
normally, aperture photometry of a SN is brighter than expected (not
fainter) if the SN is located on top of some part of the host galaxy,
because the light in the aperture is not entirely due to the SN.  That
this is not the case here must be due to the relative distributions of
host-galaxy light at the SN position and in the sky annulus.

Differences between aperture photometry and PSF photometry are 
{\em usually} larger in redder bands.  In fact, similar experiments with
adding artificial stars to the KAIT templates of SN~2001ay show that
the aperture magnitudes with the Nickel 1-m telescope are too bright
by 0.01 mag in $R$ and $I$ for the size of the aperture and annulus
used.  We never obtained near-IR subtraction templates with the camera
used on the Las Campanas Observatory 1-m telescope, and that camera
has since been decommissioned. So, we must adopt the available near-IR
aperture magnitudes.

The $I$-band photometry is particularly problematic from 15 to 30~d
after the time of $B$-band maximum.  Consider the effective filter
profiles shown in Figure \ref{i_filter}.  While the KAIT $I$-band
filter very well approximates the Bessell $I$ filter, the filter used
on the Nickel 1-m telescope does not. As the SN achieved its reddest
optical colors, a significant excess amount of light is let through by
the Nickel 1-m $I$-band filter.  This leads to positive S-corrections;
to correct such photometry to Bessell filter photometry requires
making the Nickel 1-m photometry fainter.  Our attempts to reconcile
up to 0.35 mag differences between the KAIT and Nickel 1-m photometry
were not successful.  Apparently, the effective filter profile of the
Nickel 1-m camera is even more nonstandard than our profile based on
laboratory traces of the filter, knowledge of the quantum efficiency
as a function of wavelength, and a generic atmospheric function
applicable to Lick Observatory.  While we list all of our optical
photometry in Table \ref{ubvri}, the Nickel 1-m $I$-band photometry is
not included in our plots or used in the analysis.

The $R$- and $I$-band S-corrections are shown in Figure
\ref{scorr_ri}.  Note that the photometry in Table \ref{ubvri}
includes the K- and S-corrections, plus the corrections mentioned
above for CTIO 0.9-m and Nickel 1-m photometry reduced using aperture
magnitudes.  Since the K-corrected photometry from one night with the 
YALO 1-m telescope and four nights with the LCO 2.5-m telescope are in good
agreement with photometry obtained with other telescopes, and we have
no effective transmission curves for the LCO 2.5-m filters used, we have
derived no further corrections for this small fraction of our photometry.
The interpolated $BVRI$ K-corrections and S-corrections
are listed in Table \ref{corrections}.\footnote[20]{At present,
pipelines for SN surveys use template spectra at different epochs
since maximum light, warp them, and, adopting appropriate effective
filter profiles, calculate the K- and S-corrections at the same
time.}

In spite of the rationale outlined above to reconcile as much of the
optical photometry as we can, for a derivation of the maximum
brightness and decline rate of SN~2001ay we restrict ourselves to the
photometry based on PSF magnitudes with image subtraction.  After
subtracting the derived K-corrections and adding the S-corrections, we
scale the time since maximum light to the rest frame by dividing by $1
+ z$.  We derive a time of $B$-band maximum of JD 2,452,022.49 (= 2001
April 23.0), with an uncertainty of perhaps $\pm$ 0.8~d.  The
decline-rate parameter found from our best $B$-band data is \dmm =
0.68 $\pm$ 0.05 mag.  For comparison, SNe~2005eq had \dmm\ = 0.72
$\pm$ 0.02 mag \citep{Fol_etal10}, though \citet{Con_etal10} give \dmm
= 0.78 $\pm$ 0.01 mag.  SN~2009dc had \dmm = 0.72 $\pm$ 0.03 mag
\citep{Sil_etal10}.  SNe 2001ay, 2005eq, 2006gz, 2007if, and 2009dc
had extremely slow decline rates.  Since it is well established that
the energy budget of a Type Ia SN is related to the decline rate, we
naturally wonder if SNe~2001ay and 2005eq are also
super-Chandrasekhar-mass candidates.

The $B$- and $V$-band light curves of SNe~2001ay, 2005eq, and 2009dc
are illustrated in Figure \ref{bv_comp}.  For the first three weeks
after maximum light SNe~2001ay and 2009dc were remarkably similar, but
at $t \approx 25$--40~d the light curves diverge.  As the peak-to-tail
ratio in the $B$- and $V$-band light curves sheds light on the central
density and progenitor mass of single-degenerate explosions
\citep{Hoe_etal10}, a comparison of the observational features and
modeling of very slow decliners may yield similar insights.

In Figure \ref{01ay_bv} we show the unreddened $B-V$ colors of
SN~2001ay, the unreddened locus of \citet{Lir95} and
\citet{Phi_etal99}, and the \dmm\ = 0.83 mag locus of
\citet{Pri_etal06b}.  On the basis of some of the spectra 
published here, \citet{Bra_etal06} classified SN~2001ay as a
``broad-line'' Type Ia SN.  This is the subtype that
\citet{Wan_etal09b} suggest is intrinsically redder than normal Type
Ia SNe, or occurs in dustier environments, with $R_V \approx 1.7$.
However, Figure \ref{01ay_bv} shows that, if anything, SN~2001ay is
bluer than other normal Type Ia SNe.
  
In Figure \ref{ri_comp} we show the corresponding $R$- and $I$-band
light curves of SN~2001ay, along with loci derived from the data of
SNe~2005eq and 2009dc.  The behavior of SN~2001ay in these photometric
bands is clearly more like that of SN~2009dc than of SN~2005eq.  Note
that the $I$-band secondary maximum of SN~2001ay is essentially as
bright as the first maximum.

Figure \ref{jh_comp} shows the near-IR photometry of SN~2001ay and the
other slow decliners, SNe~2005eq \citep{Fol_etal10,Con_etal10} and
2009dc \citep{Str_etal11}.  SN~1999aw \citep{Stro_etal02} and SN~2001ay
were the first objects known to exhibit such flat $H$-band light curves.
(Both $H$-band light curves are admittedly somewhat ragged, however.)
SN~2005eq was the first
to show such early near-IR maxima ($t \leq -7$~d).  Interestingly, the
$H$-band brightness of SN~2009dc increased steadily over the time
frame that the brightness of SN~2001ay was constant.

\section{Spectroscopy}  

We obtained spectra of SN~2001ay on 20 dates using 9 different
telescopes; 7 of those nights had spectra with more than one
telescope.  Spectra were obtained with the 1.5-m telescope of the Fred
L. Whipple Observatory (FLWO) on 11 nights.  Eight spectra were taken
on four nights with the MMT 6.5-m telescope; four with the 2.16-m
telescope at the Xinglong Station of the Beijing Astronomical
Observatory (BAO); three with the Las Campanas Observatory 2.5-m du
Pont telescope; two each with the Lick 3-m Shane reflector (Kast
spectrograph) and {\it HST}; and one each with the Kitt Peak 2.1-m,
Kitt Peak 4-m, and Keck II telescopes.  The spectroscopic observations
with the FLWO 1.5-m and the MMT are summarized in Table
\ref{sp1_table}; we may refer to this as the ``CfA set.''  A log of
other spectroscopic observations is given in Table
\ref{sp2_table}. Some, but not all, of the data were obtained at (or
close to) the parallactic angle \citep{Fil82} in order to minimize the
effects of atmospheric dispersion.

In Figure \ref{spectra} we show a temporal sequence of spectra.  Here
we have combined the May 2 spectrum from {\it HST} with the April 29
spectrum from the KPNO 2.1-m telescope.  The rather noisy BAO spectra
from May 20, 25, and 30 are not illustrated, but the May 10 spectrum
from BAO fills a gap in the temporal coverage.  We show additional
spectra in Figure \ref{additional_spectra}.

Figure \ref{keck} includes a portion of our spectrum of SN~2001ay
obtained with the Keck Echellette Spectrograph and Imager
\citep[ESI;][]{She_etal02} from 2001 April 22 ($t = -1$~d).  We find
that this maximum-light spectrum exhibits S~II and Si~III at $-$9000
km s$^{-1}$.  The Si~II line (rest wavelength $\lambda_0$ = 6355~\AA)
and the Mg~II line ($\lambda_0$ = 4481~\AA) are blueshifted by 14,000
km s$^{-1}$ at this epoch.  There is some gas of these species
blueshifted as much as 20,000 km s$^{-1}$.  As shown in Figure
\ref{spectra}, by $t$ = +17~d the Si~II absorption is only blueshifted
about 10,000 km s$^{-1}$.\footnote[21]{For the rest wavelengths of
  these and other species, see Table 1 of \citet{Wan_etal06}.
  In their table, however, the rest wavelength of Mg II should read
  448.1 nm, not 447.1 nm.}

In our Keck spectrum there is a small absorption dip at an observed
wavelength of $\lambda \approx 6520$~\AA.  If this is due to gas in or
near the SN or gas in the ISM of the host galaxy, the wavelength of
this line in the frame of the host galaxy is 6315~\AA. This may be due
to C~II ($\lambda_0$ = 6580~\AA) blueshifted by 12,000 km s$^{-1}$.
However, other explanations are possible.  An absorption feature at
6520~\AA\ in our Galaxy or in the Earth's atmosphere could lead to a
misidentification, given that telluric lines were not removed
from the Keck spectrum. One possible such feature is telluric 
absorption at 6515~\AA\ due to atmospheric water vapor
\citep[][Appendix]{Mat_etal00}.  We have better evidence of C~II
absorption in SN~2001ay from the $\lambda_0$ = 4745~\AA\ line, which
is seen in the galaxy's rest frame at 4555~\AA, corresponding to an
identical blueshift of 12,000 \kms.  We see no evidence of the
corresponding C~II line with $\lambda_0$ = 7234~\AA.

One might expect the C~II to be concentrated in the outer layers, and
therefore at a higher velocity of approach than sulfur or silicon
\citep{Kho_etal93, Gam_etal03}.  But if there is mixing or clumpiness,
most species could be observed over a broad range of velocities.

In Figure \ref{sodium} we see the two components of Na~I~D from gas in
the Milky Way and gas in the host galaxy of SN~2001ay.  The presence
of these lines implies some nonzero amount of reddening and
extinction.  We find equivalent widths of 0.090 $\pm$ 0.018~\AA\ for
the Galactic sodium lines and 0.210 $\pm$ 0.018 \AA\ for the
host-galaxy sodium lines.  Using the \citet{Mun_Zwi97} calibration of
the equivalent width of the sodium lines with $B-V$ color excess, we
find $E(B-V)_{\rm Gal} = 0.026 \pm 0.006$ mag and $E(B-V)_{\rm host} =
0.072 \pm 0.008$ mag.  The \citet{Mun_Zwi97} calibration shows a
scatter of $\pm 0.05$ mag for $E(B-V)$, which is a more realistic
estimate of the uncertainty of the reddening toward SN~2001ay.
\footnote[22]{However, \citet{Poz_etal11} recently found that the 
uncertainty of the reddening derived from the Na~I~D lines may be
substantially larger than this.} The Galactic component may be 
compared to $E(B-V) = 0.019$ mag obtained by \citet{Sch_etal98}.

In Figure \ref{silicon} we see the (blueshifted) velocities of Si~II
$\lambda$4130 and $\lambda$6355, with respective error bars of 
$\pm 420$ and $\pm 330$ \kms. The $\lambda$6355 line exhibits the
larger velocities because its opacity is greater; we are measuring
material farther out in the expanding fireball. The high velocities 
suggest that the outflow of SN~2001ay is essentially unimpeded, which 
is consistent with it having been a single-degenerate explosion; 
however, \citet{Mae_etal10} emphasize the importance of the viewing 
angle on the observed properties of Type Ia SNe.  We cannot state 
with certainty whether SN~2001ay was a single or double-degenerate 
explosion.

   The $\lambda$4130 line gives a velocity gradient of $\dot{v}$ =
226 $\pm$ 32 \kms\ d$^{-1}$, while the $\lambda$6355 line gives
$\dot{v}$ = 171 $\pm$ 35 \kms\ d$^{-1}$. The Si~II velocity
gradient is considerably higher than 70 \kms\ d$^{-1}$, the
criterion of \citet{Ben_etal05} to include SN~2001ay with other
``high velocity gradient'' Type Ia SNe.

\section{Discussion}

\subsection{Reddening}

As mentioned above, the Galactic reddening along the line of sight to
SN~2001ay is $E(B-V) = 0.026$ mag, and the host-galaxy component is
$E(B-V) = 0.072$ mag. Thus, for SN~2001ay, $E(B-V)_{\rm total} \approx
0.098$ mag.  Normal Galactic dust is characterized by an average value
of $R_V = A_V/E(B-V) = 3.1$ \citep {CCM89}, but dust associated with
Type Ia SNe is often characterized by a lower value; see, for example,
\citet{Kri_etal07}.  If we adopt $R_V = 3.1$ for the Galactic
component of reddening for SN~2001ay and $R_V = 2.4 \pm 0.2$ for the
host-galaxy reddening, it follows that $A_B = 0.35 \pm 0.08$, $A_V =
0.25 \pm 0.06$, $A_R = 0.21 \pm 0.05$, $A_I = 0.15 \pm 0.04$, $A_J =
0.08 \pm 0.02$, and $A_H = 0.04 \pm 0.01$ mag \citep[][Table
  8]{Kri_etal06}.

\subsection{Absolute Magnitudes at Maximum Light}

\citet{Gar_etal04} give $BVI$ decline-rate relations for what was then
the known range of \dmm\ for Type Ia SNe.  The slowest decliner used
by \citet{Pri_etal06b} for their light-curve fitting template
algorithm was SN~1999aa, with \dmm = 0.83 mag.  For such a
decline-rate parameter, the implied absolute magnitudes at maximum are
$M_B = -19.42$, $M_V = -19.39$, and $M_I = -18.85$ on an $H_0$ = 72 km
s$^{-1}$ Mpc$^{-1}$ distance scale.  Three of the four
super-Chandrasekhar-mass Type Ia SN candidates discussed by
\citet{Sca_etal10} are significantly more luminous than this.

In Figure \ref{slow_decliners} we show the absolute $V$-band
magnitudes at maximum light of SNe~2001ay, 2003fg, 2005eq, 2006gz,
2007if, and 2009dc, along with the $V$-band decline-rate relation of
\citet{Gar_etal04}. Figure \ref{absmag_ir} gives the near-IR absolute
magnitudes of Type Ia SNe at their respective maxima, including
SNe~2001ay and 2005eq.  Three of the four super-Chandra candidates are
overluminous, while all other very slow decliners have ``normal''
maximum brightness.

Based on our best optical photometry (using PSF magnitudes and
image-subtraction templates), we find apparent magnitudes for
SN~2001ay of $B_{\rm max}$ = 16.71, $V_{\rm max}$ = 16.64, $R_{\rm
  max}$ = 16.65, and $I_{\rm max}$ = 16.79.  These values include the
K-corrections and S-corrections.  Given the extinction values listed
above and the distance modulus of the host galaxy given in Table
\ref{properties}, at maximum light we find $M_B = -19.19 \pm 0.12$,
$M_V = -19.17$, $M_R = -19.10$, and $M_I = -18.90$ mag (with
uncertainties of $\pm 0.10$ mag); thus, SN~2001ay was {\em not}
overluminous at optical wavelengths.

Typically, Type Ia SNe peak in the near-IR $\sim 3$~d prior to the
time of $B$-band maximum \citep{Kri_etal04b}. SN~2005eq peaked at
least 7~d prior to $B$ maximum. We surmise that SN~2001ay may have
been $\sim 0.12$ mag brighter at maximum light than our first $J_s$
measurement, or $J_s$(max) $\approx$ 16.85 mag.  In the $H$ band 
SN~2001ay exhibited a very flat light curve; we adopt
$H_{\rm max} = 16.97 \pm 0.08$ mag.  The corresponding
extinction-corrected absolute magnitudes at IR maximum are $M_{J_s} =
-18.77 \pm 0.10$ and $M_H = -18.62 \pm 0.10$. These values may be
compared to the mean values of all but the late-peaking fast decliners
from \citet{Kri_etal09}, namely $\langle M_J \rangle = -18.61$ and
$\langle M_H \rangle = -18.31$ mag.  The standard deviations of the
distributions of the near-IR absolute magnitudes are about $\pm 0.15$
mag.  In the $J_s$ and $H$ bands, SN~2001ay was overluminous by $\sim
1\sigma$ and $2\sigma$, respectively; hence, SN~2001ay was not
statistically significantly brighter in the near-IR than other Type Ia
SNe.

\subsection{Bolometric Light Curve and Mass Budget}

In Figure \ref{bolom_lc} we show the bolometric light curve of
SN~2001ay, based on our broad-band photometry.  Though Type Ia SNe
near maximum light have spectral energy distributions that peak at
optical wavelengths, they also emit UV and IR light.  One might scale
the integrated flux by a factor of $\sim 1.1$ to account for UV and IR
light not included in the optical bandpasses
\citep{Str_etal06}, but according to \citet{Bli_etal06} this still
underestimates the bolometric flux.  Adopting distance modulus
35.55 mag ($D \approx$ 129~Mpc; see Table \ref{properties}) and a scale factor of
1.15, we obtain a peak derived bolometric luminosity ($4\pi D^2$ times
the bolometric flux) of $1.20 \times 10^{43}$ erg s$^{-1}$.  
Figure \ref{bolom_lc} also shows the bolometric light curves of SNe 2007if
and 2009dc.

At maximum light the luminosity produced by radioactive $^{56}$Ni is
given by

\begin{equation}
L_{\rm max} \; = \; \alpha \; E_{\rm Ni}~,
\end{equation}
\noindent
where $E_{\rm Ni}$ is the energy input from the decay of $^{56}$Ni,
evaluated at the time of bolometric maximum.  Arnett's Rule implies
that $\alpha$ = 1 \citep{Arn82}, meaning that the gamma-ray deposition 
matches the bolometric flux at maximum light. However, the value 
of $\alpha$ can actually range from
0.8 to 1.3, depending on the explosion model.  For a delayed
detonation model $\alpha$ = 1.2 is appropriate
\citep[][Fig. 36]{Kho_etal93}; see also \citet{Hoe_Kho96}.  
\citet{How_etal06} adopt $\alpha$ = 1.2.

In Figure \ref{edep} we show the radioactive decay energy
deposition function fit to the last few points of the bolometric
light curve.  In this figure we also show the cases of
complete trapping of the $\gamma$ rays and complete $\gamma$-ray
escape.  We adopt a rise time of 18~d from explosion to
bolometric maximum, comparable to observational results of
\citet{Gar_etal07} for other Type Ia SNe. \citet{Hay_etal10} find 
an average rise time in the $B$ band of 17.38~d, with a range of 
13 to 23~d, while the $B$-band rise time determined by 
\citet{Gan_etal11} is about 18~d.  Figure \ref{edep} uses 
$\alpha$ = 1.0.\footnote[23]{The writers of this section could
not come to a consensus on the best value of $\alpha$ to adopt,
so in what follows we give the results for $\alpha$ = 1.0 and 1.2.
\citet{Bar_etal11} choose $\alpha$ = 0.9 for their SN~2001ay model.}

The energy deposited by 1~M$_\odot$ of $^{56}$Ni into the explosion of
a Type Ia SN is given by \citet{Str_Lei05} as

\begin{equation}
E_{\rm Ni}(1~{\rm M}_{\odot}) = (6.45 \times 10^{43}) e^{-t_R/8.8} +
                           (1.45 \times 10^{43}) e^{-t_R/111.3},
\end{equation}

\noindent
where $t_R$ is the rise time in days from the moment of explosion to
the time of bolometric maximum.  The $e$-folding times of $^{56}$Ni
and $^{56}$Co are 8.8 and 111.3~d, respectively.  For
$t_R$ = 18~d and $\alpha$ = 1.0, $L_{\rm max}$ = 2.07 $\times
10^{43}$ erg s$^{-1}$ M$_{\odot}^{-1}$;  for $\alpha$ = 1.2, $L_{\rm
max}$ = 2.48 $\times 10^{43}$ erg s$^{-1}$ M$_{\odot}^{-1}$.  


There was a nondetection of SN~2001ay on April
5.4,\footnote[24]{\citet{Swi_Li01} incorrectly give the date of this
nondetection as 2001 April 4.4.} which was 17.1 rest-frame days
before the time of $B$-band maximum.  The upper limit of
the brightness on that date was $R \approx 19.5$ mag, or $\sim 3$ mag
fainter than the observed $R$-band maximum.  Although we have an image
of the host galaxy at $t = -17$~d, we would need to reach a much
fainter magnitude limit to place a useful constraint on the time of the
explosion.  The lack of premaximum photometry does not allow us to
determine the rise time of SN~2001ay.

To obtain the $^{56}$Ni yield in solar masses, we simply divide the peak
luminosity (1.20 $\times$ 10$^{43}$ erg s$^{-1}$) 
by the coefficient given above for the adopted rise time
(e.g., $2.07 \times 10^{43} \times \alpha$ for $t_R$ = 18~d);
the result is 0.58/$\alpha$ M$_{\odot}$ of $^{56}$Ni. SN~2001ay certainly 
did not produce more than a Chandrasekhar mass of $^{56}$Ni.

A 5\% uncertainty in the value of $H_0$ leads to a 10\% uncertainty in
the luminosity calculated from the optical photometry.  Given the
uncertainties in the parameter $\alpha$ in Equation 1 and
uncertainty in the adopted bolometric rise time, the
minimum uncertainty in the $^{56}$Ni yield is about $\pm 0.15$~M$_{\odot}$
divided by $\alpha$.

We would like to estimate the mass in the ejecta of SN~2001ay. 
Following \citet{Jef99} and \citet{Str_etal06}, the ``fiducial time'' 
($t_0$) is the time scale for the ejecta to become optically thin:

\begin{equation}
t_0 \; = \; \left( \frac{M_{\rm ej} \kappa q}{8 \pi} \right) ^{\onehalf} \frac{1}{v_e} \; ,
\end{equation}

\parindent = 0 mm

where $M_{\rm ej}$ is the ejecta mass, $\kappa$ is the $\gamma$-ray
mean opacity, $q$ is a dimensionless scale factor, and $v_e$ is the
$e$-folding velocity of an exponential model's density profile.
\citet{Str_etal06} adopt $\kappa$ = 0.025 cm$^2$ g$^{-1}$ and $v_e$ =
3000 km s$^{-1}$. They also choose $q = 1/3$, meaning that $^{56}$Ni
was distributed throughout the ejecta; $q = 1$ for high concentrations
of $^{56}$Ni at the center of the ejecta.  Using the last three points
in our bolometric light curve (from $t$ = 49 to 111~d), adopting
$\alpha$ = 1.2, and using a $^{56}$Ni yield of 0.48 M$_{\odot}$ we
derive a ``fiducial time'' of 66.2 $\pm$ 3.0~d.  Taking $\alpha$ = 1.0 and 
a $^{56}$Ni yield of 0.58 M$_{\odot}$ gives $t_0$ = 57.2 $\pm$ 2.4 d.
Both of these values are considerably longer than the 22~d to
35~d in Table 1 of \citet{Str_etal06}, but not much
greater than the 51~d for SN~2007if given in Figure 9 of
\citet{Sca_etal10}.  The SN~2001ay fiducial time depends critically on
the final point in the bolometric light curve, a point derived from
PSF photometry without template subtraction (hence possibly erroneous).
Using $t_0$ = 66.2 d and the values of $q$, $v_e$, and $\kappa$
adopted by \citet{Str_etal06}, the implied ejecta mass is 4.4
M$_{\odot}$, which is clearly wrong.  For $t_0$ = 57.2 d and the
higher $^{56}$Ni yield, the implied ejecta mass is 3.3 M$_{\odot}$.
Adopting $q = 1$ instead could cut these down by a factor of three.
Choosing $q = 1/3$ but $v_e$ = 1500 km s$^{-1}$ would give 0.8
to 1.1 M$_{\odot}$ for the ejecta mass.  The lower $e$-folding velocity
is only slightly less than what one derives from the three-dimensional 
models of \citet{Rop_Hil04}.  Given the uncertainties of the parameters
necessary for Eq. 3 and the lack of high-quality photometry from
$\sim 50$ to 100~d after maximum light, we feel that a robust
calculation of the ejecta mass of SN~2001ay is beyond the scope of
this paper.

\parindent = 9 mm

We note that \citet{Tau_etal11} derived an ejecta mass for SN~2009dc
of 2.8~M$_{\odot}$, based on the expansion velocity and the timescale
around maximum brightness, which depends on the optical opacity.
They describe their result as ``an utmost challenge for all scenarios that
invoke thermonuclear explosions of white dwarfs.''

\subsection{Spectroscopic Comparison and Modeling}

\citet{Wag_etal10} have used wavelet analysis to search for
spectroscopic correlations of Type Ia SNe.  Their results
are based on the analysis of a few dozen relatively nearby
objects ($z \lesssim 0.04$).  SN~2001ay is anomalous
in a number of their plots, particularly their Figure 12,
showing the correlation of the strength of the 4570 \AA\
emission feature vs. \dmm.

In Figure \ref{3spectra} we illustrate a comparison of the spectra of
SNe~2001ay, 2005eq, and 2009dc.  The most obvious difference between
SN~2001ay and these other two objects is the much larger blueshift of
Si~II seen in SN~2001ay prior to $B$ maximum.  Also, the C~II
absorption in SN~2009dc is much stronger.  SN~2005eq was apparently
the hottest of the three, given the stronger presence of doubly
ionized iron and the weakness of singly ionized species.  This is more
like the classical slow decliner SN~1991T \citep{Fil97}, which is, in
fact, how \citet[][Table 1]{Con_etal10} classify SN~2005eq.

In Table \ref{three} we give a summary of some observational
characteristics of SNe~2001ay, 2005eq, and 2009dc.  Given the
divergence of the light curves of SN~2005eq seen in Figure
\ref{bv_comp} compared to the other two objects, the larger
decline-rate parameter of SN~2005eq (\dmm\ = 0.78 mag) from
\citet{Con_etal10} is more likely correct than the value of \dmm\ =
0.72 mag from \citet{Fol_etal10}, though we have used the latter to
plot SN~2005eq in Figure \ref{slow_decliners}.  SNe~2001ay and 2009dc
have almost the same optical decline rate.  SN~2001ay had normal
optical peak brightness and high Si~II velocity.  SN~2009dc was
overluminous at optical maximum and had much lower Si~II velocity.
Their $H$-band light curves are unlike those of any other Type Ia SNe
observed thus far, except for SN~1999aw.

To investigate the details of our SN spectra, we use the SN
spectrum-synthesis code SYNOW \citep{Fis_etal97}.  Although SYNOW has
a simple, parametric approach to creating synthetic spectra, it is a
powerful tool to aid line identifications which in turn provide
insights into the spectral formation of the objects. To generate a
synthetic spectrum one inputs a blackbody temperature ($T_{\rm BB}$),
a photospheric velocity ($v_{\rm ph}$), and for each involved ion, 
an optical depth at a reference line, an excitation temperature 
($T_{\rm exc}$), and the maximum velocity of the opacity distribution
($v_{\rm max}$).  Moreover, it assumes that the optical depth declines
exponentially for velocities above $v_{\rm ph}$ with an $e$-folding
scale of $v_{e}$.  The strengths of the lines for each ion are
determined by oscillator strengths, and the approximation of a
Boltzmann distribution of the lower level populations is set by the
temperature $T_{\rm exc}$.

In Figure \ref{synow1}, we present our $-1$~d spectrum of SN~2001ay
with a synthetic spectrum generated from SYNOW. This fit has $T_{\rm
  BB}$ = 20,500~K and $v_{\rm ph} = 12,000$~km~s$^{-1}$. The majority
of the observed features are well fit by the synthetic spectrum. The
ions used in the fit, as labeled in the figure, are commonly observed
in the near-maximum spectra of Type Ia SNe \citep{Bra_etal05} with the
exception of C~II. Although the inclusion of C~II in the fit produces
an absorption feature at $\sim 6300$~\AA\ (due to C~II $\lambda$6580)
that is too strong compared to the observed spectrum, it nonetheless
yields a better fit to the absorption feature at $\sim 4500$~\AA\ (due
to C~II $\lambda$4745), so we conclude that the inclusion of C~II
marginally improves the SYNOW fit.

We attempt SYNOW fits to the SN 2001ay spectra at $t \approx +6$~d and
23~d.  Figure \ref{synow2} shows the result for the $t \approx +6$~d
spectrum. The model spectrum has $T_{\rm BB}$ = 14,000~K and $v_{\rm
  ph}$ = 11,000~km~s$^{-1}$, and has all the regular ions observed in
a Type Ia SN.  C~II is no longer needed, but a relatively strong Na~I
line is now observed at $\sim 5600$~\AA. A SYNOW fit for the $t
\approx 23$~d spectrum (not shown) requires $T_{\rm BB} = 8500$~K and
$v_{\rm ph} = 9000$~km~s$^{-1}$.

%

Finally, in Figure \ref{synapps} we show a model fit to our Keck
spectrum using the SYNAPPS code of \citet{Tho_etal11}.  The physical
assumptions SYNAPPS uses match those of SYNOW \citep{Fis_etal97},
so findings are restricted to identification of features and not
quantitative abundances.  But where SYNOW is completely interactive,
SYNAPPS is automated.  This relieves the user from iterative
adjustment of a large number of parameters (over 50 variables) to gain
fit agreement and assures more exhaustive searching of the parameter
space.  SYNAPPS can be thought of as the hybridization of a SYNOW-like
calculation with a parallel optimization framework, where spectral fit
quality serves as the objective function to optimize.  A good fit
constrains explosion models through interpretive spectral feature
identification, with the main result being the detection or exclusion
of specific chemical elements.  The velocity distribution of detected
species within the ejecta can also be constrained.

SYNAPPS indicates that SN~2001ay had a photospheric velocity of
10,800 \kms\ at $t = -1$ d, which should be robust, as it is derived 
from 11 ions, namely C~II, O~I, Mg~II, Si~II, Si~III, S~II, Ca~II,
Fe~II, Fe~III, Co~II, and Ni~II.  We are confident of the presence 
of C~II.

\section{Conclusions}

We have presented the available spectra, as well as optical and
near-IR photometry, of SN~2001ay.

%

While SN~2001ay is the most slowly declining Type Ia SN ever
discovered, it was not overluminous in optical bandpasses.  In near-IR
bands it was overluminous only at the 2$\sigma$ level at most.
Unlike other very slow decliners such as SNe 2003fg, 2007if, and
2009dc, which were significantly overluminous, we do not have any
evidence that SN~2001ay was a super-Chandrasekhar-mass explosion.

SN~2001ay showed evidence for C~II, but it was much weaker than in
SN~2009dc.  At early times Mg~II and Si~II were observed in SN~ 2001ay
at high velocity (14,000 \kms\ and higher).  By contrast, spectra of
SN 2009dc did not show large velocities for Mg~II and Si~II.  On the
basis of a small number of super-Chandrasekhar-mass candidates, it
seems that these objects exhibit rather low velocities, possibly the
result of retardation due to a shell of material arising from the
disruption of the less massive white dwarf in a double-degenerate
system.

SN~2001ay produced (0.58 $\pm$ 0.15)/$\alpha$ M$_{\odot}$ of $^{56}$Ni,
considerably less than a Chandrasekhar mass. The value of $\alpha$
probably lies in the range 1.0 to 1.2.  Naively, one might conclude
that SN~2001ay was a single-degenerate explosion, but this is hardly a
firm conclusion.  The very broad light curve might be explained by the
trapping of the $\gamma$ rays in the explosion.  An explanation of the
extremely slow decline will be discussed in a subsequent paper
\citep{Bar_etal11}.

\acknowledgments

The work presented here is based in part on observations made with the
NASA/ESA {\it Hubble Space Telescope}, obtained at the Space Telescope
Science Institute, which is operated by the Association of
Universities for Research in Astronomy, Inc., under NASA contract
NAS5-26555; the Cerro Tololo Inter-American Observatory and the Kitt
Peak National Observatory of the National Optical Astronomy
Observatory, which is operated by the Association of Universities for
Research in Astronomy, Inc. (AURA) under cooperative agreement with
the NSF; the MMT Observatory, a joint facility of the Smithsonian
Institution and the University of Arizona; the Fred L. Whipple
Observatory; the Lick Observatory of the University of California; the
Las Campanas Observatory; the Beijing Astronomical Observatory; and
the W.~M. Keck Observatory, which was generously funded by the
W.~M. Keck Foundation and is operated as a scientific partnership
among the California Institute of Technology, the University of
California, and NASA. We thank the staffs at these observatories for
their efficient assistance, Don Groom for taking some of the Nickel
1-m images, and Rachel Gibbons, Maryam Modjaz, Isobel Hook, and
Saul Perlmutter for other
observational assistance. We are grateful to Peter H\"{o}flich, Alexei
Khokhlov, and Eddie Baron for comments on \S 4.3.

The supernova research of A.V.F.'s group at U.~C. Berkeley is
supported by NSF grant AST-0908886 and by the TABASGO Foundation, as
well as by NASA through grants AR-11248 and AR-12126 from the Space
Telescope Science Institute, which is operated by Associated Universities for
Research in Astronomy, Inc., under NASA contract NAS 5-26555.
KAIT and its ongoing operation were made possible by donations from
Sun Microsystems, Inc., the Hewlett-Packard Company, AutoScope
Corporation, Lick Observatory, the NSF, the University of California,
the Sylvia \& Jim Katzman Foundation, and the TABASGO Foundation.
J.M.S. is grateful to Marc J. Staley for a Graduate Fellowship.
K. K., L. W., and N. B. S. are supported in part by NSF grant AST-0709181.
Supernova research at Harvard is supported by NSF
grant AST-0907903. This work was also supported by the
Director, Office of Science, Office of High Energy Physics, of the U.S.
Department of Energy under Contract No. DE-AC02-05CH11231.

\clearpage

\begin{deluxetable}{ll}
\tablewidth{0pt}
\tablecolumns{2}
\tablecaption{Properties of SN~2001ay and its Host Galaxy\label{properties}}
\tablehead{
\colhead{Parameter} & \colhead{Value}
}
\startdata
Host galaxy                                    & IC 4423 \\
Host-galaxy type\tablenotemark{a}              & Sbc      \\
Heliocentric radial velocity\tablenotemark{b}  & 9067\,\kms  \\
CMB-frame radial velocity\tablenotemark{c}     & 9266\,\kms  \\
Distance modulus\tablenotemark{d}              & $35.55 \pm 0.1$ mag          \\
$E(B-V)_{\rm Gal}$                             & 0.026 $\pm$ 0.006 mag \\
$E(B-V)_{\rm host}$                            & 0.072 $\pm$ 0.008 mag \\
SN $\alpha$(J2000)   & $14^{\rm h}26^{\rm m}17^{\rm s}\hspace{-1 mm}.0$  \\
SN $\delta$(J2000)                             & $+26\degr 14'55\farcs8$  \\
Offset from nucleus                            & 10\farcs3\,W \, 9\farcs3\,N  \\
Julian Date of $B$-band maximum                & 2,452,022.49 $\pm 0.8$ \\
UT Date of $B$-band maximum                    & 2001 April 23.0 \\
$\Delta m_{15}(B)$                             & 0.68 $\pm$ 0.05 mag    \\
$M_B$(max)                                     & $-19.19 \pm 0.12$ \\
$M_V$(max)                                     & $-19.17 \pm 0.10$ \\
$M_R$(max)                                     & $-19.10 \pm 0.10$ \\
$M_I$(max)                                     & $-18.90 \pm 0.10$ \\
$M_J$(max)                                     & $-18.77 \pm 0.10$ \\
$M_H$(max)                                     & $-18.62 \pm 0.10$ \\
\enddata
\tablenotetext{a}{From HyperLEDA.}
\tablenotetext{b}{\citet{Bee_etal95}, via NED.}
\tablenotetext{c}{From NED.}
\tablenotetext{d}{Using $H_0$ = 72 \kms\ Mpc$^{-1}$ \citep{Fre_etal01}.}
\end{deluxetable}

\begin{deluxetable}{rccccc}
\tablewidth{0pt}
\tablecolumns{6}
\tablecaption{Optical Field-Star Sequence near SN~2001ay\tablenotemark{a}\label{ubvri_field}}
\tablehead{
\colhead{ID\tablenotemark{b}}  &
\colhead{$U$ (mag)} &
\colhead{$B$ (mag)} &
\colhead{$V$ (mag)} &
\colhead{$R$ (mag)} &
\colhead{$I$ (mag)}
}
\startdata
1 & 17.112 (0.043) &  16.399 (0.019) & 15.480 (0.008) &  14.859 (0.018) &  14.386 (0.015) \\
2 & 18.224 (0.047) &  17.262 (0.020) & 16.193 (0.007) &  15.467 (0.019) &  14.921 (0.016) \\
3 & 16.705 (0.022) &  16.846 (0.019) & 16.367 (0.008) &  15.981 (0.022) &  15.622 (0.021) \\
4 & 17.933 (0.035) &  18.271 (0.030) & 17.864 (0.005) &  17.503 (0.025) &  17.131 (0.025) \\
5 & 19.246 (0.174) &  18.729 (0.027) & 17.884 (0.015) &  17.302 (0.024) &  16.796 (0.033) \\
6 & 16.813 (0.031) &  16.853 (0.017) & 16.320 (0.007) &  15.917 (0.017) &  15.574 (0.014) \\
7 & 20.437 (0.129) &  19.585 (0.015) & 18.313 (0.010) &  17.487 (0.017) &  16.730 (0.010) \\
8 & 20.137 (0.230) &  19.987 (0.196) & 18.674 (0.070) &  18.110 (0.100) &  17.467 (0.111) \\
9 & 21.270 (0.289) &  19.802 (0.108) & 18.790 (0.059) &  18.009 (0.101) &  17.467 (0.074) \\
10 &  \ldots       &    \ldots       &   \ldots       &  20.719 (0.155) &  18.930 (0.061) \\
\enddata
\tablenotetext{a}{Magnitude uncertainties ($1\sigma$) are given 
in parentheses.}
\tablenotetext{b} {The IDs are the same as in Fig. \ref{finder}.  
Star 10, located $18.4''$ SE of Star 7, is not visible in the 
$V$-band finder.}
\end{deluxetable}

\begin{deluxetable}{rcc}
\tablewidth{0pt}
\tablecolumns{6}
\tablecaption{Infrared Field Star Sequence near SN~2001ay\tablenotemark{a}\label{jh_field}}
\tablehead{
\colhead{ID\tablenotemark{b}}  &
\colhead{$J_s$ (mag)} &
\colhead{$H$ (mag)} 
}
\startdata
1 & 13.718 (0.014) & 13.251 (0.011) \\
6 & 15.153 (0.021) & 14.881 (0.008) \\
7 & 15.869 (0.007) & 15.281 (0.007) \\
8 & 16.566 (0.030) & 15.932 (0.031) \\
10 & 17.838 (0.024) & 17.359 (0.029) \\
\enddata
\tablenotetext{a}{Magnitude uncertainties ($1\sigma$) are given 
in parentheses.}
\tablenotetext{b}{The IDs are the same as in Fig. \ref{finder} and 
Table \ref{ubvri_field}.}
\end{deluxetable}

\begin{deluxetable}{lccccccc}
\tablewidth{0pt}
\tabletypesize{\scriptsize}
\tablecolumns{8}
\tablecaption{Fully Corrected Optical Photometry of SN~2001ay\tablenotemark{a}\label{ubvri}}
\tablehead{
\colhead{JD\tablenotemark{b}} &
\colhead{$U$ (mag)} &
\colhead{$B$ (mag)} &
\colhead{$V$ (mag)} &
\colhead{$R$ (mag)} &
\colhead{$I$ (mag)} &
\colhead{Telescope\tablenotemark{c}} &
\colhead{Subs\tablenotemark{d}}
}
\startdata
2020.78 & \ldots         & 16.773 (0.043) & 16.708 (0.010) & 16.694 (0.015) & 16.771 (0.017) & 1 & N \\
2021.71 & 16.419 (0.034) & 16.731 (0.020) & 16.700 (0.010) & 16.695 (0.006) & 16.802 (0.010) & 1 & N \\ 
\hline
2022.90 &  \ldots        & 16.687 (0.050) & 16.643 (0.020) & 16.635 (0.040) & 16.783 (0.050) & 2 & Y \\
2023.88 &  \ldots        & 16.724 (0.070) & 16.640 (0.020) & 16.661 (0.030) & 16.781 (0.020) & 2 & Y \\
2024.89 &  \ldots        & 16.709 (0.040) & 16.647 (0.020) & 16.626 (0.020) & 16.829 (0.030) & 2 & Y \\
2025.89 &  \ldots        & 16.737 (0.030) & 16.666 (0.020) & 16.692 (0.020) & 16.868 (0.050) & 2 & Y \\
2026.86 &  \ldots        & 16.766 (0.020) & 16.664 (0.020) & 16.659 (0.030) & 16.855 (0.040) & 2 & Y \\
2028.89 &  \ldots        & 16.841 (0.100) & 16.699 (0.080) & \ldots         & \ldots         & 2 & Y \\
2030.88 &  \ldots        & 16.913 (0.040) & 16.734 (0.020) & 16.775 (0.030) & 16.937 (0.040) & 2 & Y \\
2032.88 &  \ldots        & 17.065 (0.050) & 16.799 (0.030) & 16.860 (0.030) & 16.952 (0.050) & 2 & Y \\
2037.88 &  \ldots        & 17.290 (0.050) & 16.939 (0.030) & 16.948 (0.020) & 17.102 (0.040) & 2 & Y \\
2039.86 &  \ldots        & 17.535 (0.080) & 16.945 (0.050) & 16.963 (0.040) & 17.057 (0.070) & 2 & Y \\
2041.80 &  \ldots        & 17.647 (0.040) & 17.089 (0.050) & \ldots         & \ldots         & 2 & Y \\
2043.85 &  \ldots        & \ldots         & 17.143 (0.020) & 16.968 (0.020) & 17.093 (0.050) & 2 & Y \\
2045.86 &  \ldots        & 17.952 (0.050) & 17.266 (0.030) & 17.006 (0.020) & 17.024 (0.040) & 2 & Y \\
2047.84 &  \ldots        & 18.134 (0.040) & 17.298 (0.040) & 17.015 (0.030) & 16.969 (0.040) & 2 & Y \\
2051.81 &  \ldots        & 18.250 (0.050) & 17.424 (0.030) & 17.123 (0.020) & 16.902 (0.040) & 2 & Y \\
2059.81 &  \ldots        & 18.554 (0.080) & 17.794 (0.040) & 17.385 (0.020) & 17.138 (0.040) & 2 & Y \\
2064.83 &  \ldots        & 18.773 (0.080) & 18.064 (0.070) & 17.629 (0.040) & 17.251 (0.060) & 2 & Y \\
2069.79 &  \ldots        & 19.025 (0.070) & 18.117 (0.060) & 17.813 (0.030) & 17.506 (0.060) & 2 & Y \\
2074.74 &  \ldots        & 19.071 (0.070) & 18.219 (0.070) & 18.008 (0.030) & 17.805 (0.060) & 2 & Y \\
2075.81 &  \ldots        & 18.993 (0.060) & 18.153 (0.060) & 18.099 (0.050) & 17.604 (0.050) & 2 & Y \\
2079.75 &  \ldots        & \ldots         & 18.250 (0.070) & 18.180 (0.050) & 17.804 (0.060) & 2 & Y \\
\hline
2022.99 &  \ldots        & 16.709 (0.033) & 16.655 (0.013) & 16.693 (0.011) & 16.758 (0.016) & 3 & Y \\ 
2024.91 &  \ldots        & 16.795 (0.044) & 16.629 (0.025) & 16.686 (0.011) & 16.781 (0.027) & 3 & Y \\
2029.81 &  \ldots        & 16.958 (0.060) & 16.716 (0.010) & 16.781 (0.010) & 16.910 (0.015) & 3 & Y \\
2036.86 & \ldots         & 17.322 (0.020) & 16.932 (0.010) & 16.901 (0.023) & 17.007 (0.024) & 3 & N \\
2037.83 & 17.503 (0.040) & 17.348 (0.019) & 16.990 (0.010) & 16.967 (0.018) & 17.099 (0.020) & 3 & N \\
2038.77 & \ldots         & 17.488 (0.009) &   \ldots       & 17.043 (0.010) & 17.079 (0.010) & 3 & N \\
2040.79 & 17.655 (0.031) & 17.578 (0.014) & 17.054 (0.010) & 16.964 (0.016) & 16.939 (0.016) & 3 & N \\
2041.84 & \ldots         & 17.668 (0.034) & 17.003 (0.020) & 16.949 (0.033) & 16.910 (0.034) & 3 & N \\
2043.78 & \ldots         & 17.784 (0.010) & \ldots         & 16.990 (0.010) & 16.845 (0.017) & 3 & N \\
2046.82 & \ldots         &  \ldots        & \ldots         & 17.107 (0.010) & 16.871 (0.010) & 3 & N \\
2050.82 & \ldots         & 18.269 (0.013) & \ldots         & 16.992 (0.085) & 16.816 (0.010) & 3 & N \\
2063.76 & \ldots         & \ldots         & 18.083 (0.017) & \ldots         & 17.225 (0.029) & 3 & N \\
2140.70 &  \ldots        & 20.110 (0.130) & 19.520 (0.100) & 19.390 (0.080) & 19.030 (0.090) & 3 & N \\
\hline
2026.74 &  \ldots        & 16.841 (0.062) & 16.673 (0.040) & \ldots         &  \ldots        & 4 & N \\
\hline
2030.67 & \ldots         & 16.953 (0.017) & 16.725 (0.010) &  \ldots        &  \ldots        & 5 & N \\ 
2046.62 & \ldots         & 18.063 (0.024) & 17.194 (0.013) &  \ldots        &  \ldots        & 5 & N \\   
2047.65 & \ldots         & 18.209 (0.047) & 17.247 (0.027) & \ldots         &  \ldots        & 5 & N \\
2055.63 & \ldots         & \ldots         & 17.687 (0.028) & \ldots         & \ldots         & 5 & N \\
\hline
2052.65 & 18.633 (0.041) & 18.356 (0.023) & 17.468 (0.012) & 17.152 (0.027) & 16.850 (0.029) & 6 & Y \\
\enddata

\tablenotetext{a}{The $BVRI$ data in this table include the K-corrections, 
S-corrections, and corrections applied to photometry derived using aperture 
magnitudes without subtraction templates. Magnitude uncertainties ($1\sigma$) 
are given in parentheses.}
\tablenotetext{b}{Julian Date {\em minus} 2,450,000.}
\tablenotetext{c}{1 = CTIO 0.9-m, 2 = KAIT, 3 = Nickel 1-m, 4 = Yale-AURA-Lisbon-Ohio (YALO) 1-m,
5 = LCO 2.5-m, and 6 = CTIO 1.5-m telescopes.}
\tablenotetext{d}{Image-subtraction templates used? Y = yes; N = No. }
\end{deluxetable}


\begin{deluxetable}{cclcccc}
\tablecolumns{7}
\tablewidth{0pc}
\tabletypesize{\scriptsize}
\tablecaption{K-Corrections for SN 2001ay in Optical Bands\tablenotemark{a}\label{opt_kcorrs}}
\tablehead{
\colhead{UT Date (2001)}  &
\colhead{Phase (d)\tablenotemark{b}} &
\colhead{Telescope} &
\colhead{$\Delta B$ (mag)} &
\colhead{$\Delta V$ (mag)} &
\colhead{$\Delta R$ (mag)} &
\colhead{$\Delta I$ (mag)}
}
\startdata
Apr 21 & \phn  $-$2 & FLWO 1.5-m &    $-$0.010 &    $-$0.031 & $-$0.120 &    $-$0.045 \\
Apr 24 & \phn    +1 & FLWO 1.5-m &  \phn 0.000 &    $-$0.028 & $-$0.118 &    $-$0.068 \\
Apr 26 & \phn    +3 & MMT 6.5-m  &    $-$0.001 &    $-$0.020 & $-$0.090 &    $-$0.029 \\
Apr 27 & \phn    +4 & FLWO 1.5-m &  \phn 0.018 &    $-$0.014 & $-$0.097 &    $-$0.015 \\
Apr 30 & \phn    +7 & FLWO 1.5-m &  \phn 0.010 &    $-$0.017 & $-$0.094 &    $-$0.050 \\
May 02 & \phn    +9 & FLWO 1.5-m &  \phn 0.023 &    $-$0.005 & $-$0.093 &    $-$0.065 \\
May 10 &        +17 & BAO 2.16-m &  \ldots     &    $-$0.024: & $-$0.064 &    \ldots  \\
May 16 &        +23 & FLWO 1.5-m &  \phn 0.080 & \phn  0.047 & $-$0.055 &  \phn 0.042 \\
May 23 &        +30 & FLWO 1.5-m &  \phn 0.090 & \phn  0.066 & $-$0.057 &  \phn 0.030 \\
May 24 &        +31 & MMT  6.5-m &  \phn 0.097 & \phn  0.077 & $-$0.039 &    $-$0.008 \\
May 25 &        +32 & FLWO 1.5-m &  \phn 0.085 & \phn  0.068 & $-$0.061 & \phn  0.025 \\
May 25 &        +32 & MMT 6.5-m  &  \phn 0.100 & \phn  0.081 & $-$0.040 & \phn  0.022 \\
May 30 &        +37 & FLWO 1.5-m &  \phn 0.082 & \phn  0.070 & $-$0.049 & \phn  0.019 \\
Jun 18 &        +56 & FLWO 1.5-m &  \phn 0.089 & \phn  0.073 & $-$0.067 & \phn  0.043 \\
Jun 18 &        +56 & MMT 6.5-m  &  \phn 0.093 & \phn  0.070 & $-$0.056 & \phn  0.059 \\
\tableline
\enddata
\tablenotetext{a} {These corrections (measured in magnitudes)
are {\em subtracted} from the photometric data to 
correct the photometry to the rest frame.  These were 
calculated using the \citet{Bes90} filter profiles.}
\tablenotetext{b} {The number of observer-frame days since
the time of $B$ maximum, 2001 April 23.}
\end{deluxetable}

\begin{deluxetable}{lccccccccc}
\tablewidth{0pt}
\tabletypesize{\scriptsize}
\tablecolumns{8}
\tablecaption{Interpolated Corrections to Optical Photometry\tablenotemark{a}\label{corrections}}
\tablehead{
\colhead{JD\tablenotemark{b}} &
\colhead{K$_{C_B}$} &
\colhead{K$_{C_V}$} &
\colhead{K$_{C_R}$} &
\colhead{K$_{C_I}$} &
\colhead{S$_{C_B}$} &
\colhead{S$_{C_V}$} &
\colhead{S$_{C_R}$} &
\colhead{S$_{C_I}$} &
\colhead{Telescope\tablenotemark{c}}
}
\startdata
2020.78 &   $-$0.010 &   $-$0.031 & $-$0.120 &   $-$0.053 & $-$0.008 &   $-$0.003 &   $-$0.006 & $-$0.002 & 1 \\ 
2021.71 &   $-$0.006 &   $-$0.030 & $-$0.116 &   $-$0.052 & $-$0.007 &   $-$0.004 &   $-$0.005 & $-$0.006 & 1 \\
\hline
2022.90 &   $-$0.005 &   $-$0.028 & $-$0.113 &   $-$0.050 & $-$0.008 &   $-$0.005 & \phn 0.022 & $-$0.017 & 2 \\
2023.88 &   $-$0.002 &   $-$0.025 & $-$0.110 &   $-$0.048 & $-$0.008 &   $-$0.005 & \phn 0.021 & $-$0.017 & 2 \\
2024.89 & \phn 0.004 &   $-$0.022 & $-$0.106 &   $-$0.046 & $-$0.007 &   $-$0.005 & \phn 0.020 & $-$0.017 & 2 \\
2025.89 & \phn 0.006 &   $-$0.021 & $-$0.103 &   $-$0.044 & $-$0.007 &   $-$0.005 & \phn 0.019 & $-$0.016 & 2 \\
2026.86 & \phn 0.007 &   $-$0.019 & $-$0.100 &   $-$0.041 & $-$0.007 &   $-$0.005 & \phn 0.019 & $-$0.016 & 2 \\

2028.89 & \phn 0.013 &   $-$0.013 & \ldots   & \ldots     & $-$0.006 &   $-$0.004 & \ldots     & \ldots   & 2 \\
2030.88 & \phn 0.022 &   $-$0.006 & $-$0.088 &   $-$0.033 & $-$0.005 &   $-$0.002 & \phn 0.017 & $-$0.016 & 2 \\
2032.88 & \phn 0.030 & \phn 0.000 & $-$0.083 &   $-$0.028 & $-$0.005 &   $-$0.001 & \phn 0.017 & $-$0.016 & 2 \\
2037.88 & \phn 0.045 & \phn 0.016 & $-$0.071 &   $-$0.017 & $-$0.005 & \phn 0.005 & \phn 0.017 & $-$0.015 & 2 \\
2039.86 & \phn 0.050 & \phn 0.022 & $-$0.066 &   $-$0.012 & $-$0.005 & \phn 0.007 & \phn 0.017 & $-$0.015 & 2 \\
 
2041.80 & \phn 0.057 & \phn 0.029 & \ldots   & \ldots     & $-$0.006 & \phn 0.008 & \ldots     & \ldots   & 2 \\
2043.85 & \ldots     & \phn 0.037 & $-$0.059 &   $-$0.007 & \ldots   & \phn 0.010 & \phn 0.019 & $-$0.014 & 2 \\
2045.86 & \phn 0.071 & \phn 0.045 & $-$0.056 & \phn 0.002 & $-$0.007 & \phn 0.011 & \phn 0.020 & $-$0.014 & 2 \\
2047.84 & \phn 0.078 & \phn 0.053 & $-$0.054 & \phn 0.007 & $-$0.008 & \phn 0.011 & \phn 0.021 & $-$0.014 & 2 \\
2051.81 & \phn 0.089 & \phn 0.068 & $-$0.050 & \phn 0.015 & $-$0.011 & \phn 0.012 & \phn 0.023 & $-$0.013 & 2 \\

2059.81 & \phn 0.089 & \phn 0.075 & $-$0.048 & \phn 0.031 & $-$0.017 & \phn 0.009 & \phn 0.027 & $-$0.011 & 2 \\
2064.83 & \phn 0.089 & \phn 0.073 & $-$0.050 & \phn 0.039 & $-$0.018 & \phn 0.007 & \phn 0.029 & $-$0.010 & 2 \\
2069.79 & \phn 0.087 & \phn 0.067 & $-$0.053 & \phn 0.045 & $-$0.018 & \phn 0.004 & \phn 0.030 & $-$0.009 & 2 \\
2074.74 & \phn 0.088 & \phn 0.066 & $-$0.058 & \phn 0.048 & $-$0.011 & \phn 0.005 & \phn 0.030 & $-$0.007 & 2 \\
2075.81 & \phn 0.088 & \phn 0.067 & $-$0.059 & \phn 0.049 & $-$0.009 & \phn 0.005 & \phn 0.030 & $-$0.007 & 2 \\
2079.75 & \ldots     & \phn 0.072 & $-$0.062 & \phn 0.050 & \ldots   & \phn 0.010 & \phn 0.028 & $-$0.006 & 2 \\
\hline
2022.99 &   $-$0.005 &   $-$0.028 & $-$0.113 &   $-$0.050 & \phn 0.003 &   $-$0.009 & \phn 0.024 & \phn 0.027 & 3 \\
2024.91 & \phn 0.004 &   $-$0.022 & $-$0.106 &   $-$0.046 & \phn 0.004 &   $-$0.008 & \phn 0.023 & \phn 0.036 & 3 \\
2029.81 & \phn 0.017 &   $-$0.010 & $-$0.091 &   $-$0.035 & \phn 0.000 &   $-$0.002 & \phn 0.022 & \phn 0.060 & 3 \\
 
2036.86 & \phn 0.042 & \phn 0.013 & $-$0.073 &   $-$0.019 &   $-$0.014 & \phn 0.007 & \phn 0.024 & \phn 0.073 & 3 \\
2037.83 & \phn 0.045 & \phn 0.016 & $-$0.071 &   $-$0.017 &   $-$0.016 & \phn 0.008 & \phn 0.024 & \phn 0.076 & 3 \\
2038.77 & \phn 0.047 & \ldots     & $-$0.068 &   $-$0.014 &   $-$0.018 & \ldots     & \phn 0.024 & \phn 0.079 & 3 \\
2040.79 & \phn 0.053 & \phn 0.026 & $-$0.064 &   $-$0.010 &   $-$0.022 & \phn 0.011 & \phn 0.024 & \phn 0.082 & 3 \\
2041.84 & \phn 0.057 & \phn 0.029 & $-$0.063 &   $-$0.007 &   $-$0.024 & \phn 0.011 & \phn 0.024 & \phn 0.082 & 3 \\
2043.78 & \phn 0.064 & \ldots     & $-$0.059 &   $-$0.003 &   $-$0.028 & \ldots     & \phn 0.024 & \phn 0.082 & 3 \\
2046.82 &   \ldots   & \ldots     & $-$0.056 & \phn 0.004 & \ldots     & \ldots     & \phn 0.025 & \phn 0.075 & 3 \\
2050.82 & \phn 0.087 & \ldots     & $-$0.051 & \phn 0.013 &   $-$0.037 & \ldots     & \phn 0.027 & \phn 0.053 & 3 \\
2063.76 &  \ldots    & \phn 0.074 & \ldots   & \phn 0.037 &   \ldots   &   $-$0.005 & \ldots     & \phn 0.000 & 3 \\
\hline
2026.74 & \phn 0.007 &   $-$0.019 & \ldots    &  \ldots    &   \ldots  &    \ldots  & \ldots    & \ldots       & 4 \\
\hline 
2030.67 & \phn 0.008 &   $-$0.006 &  \ldots   &  \ldots    &   \ldots  &   \ldots   & \ldots    & \ldots       & 5 \\
2046.62 & \phn 0.075 & \phn 0.049 &  \ldots   &  \ldots    &   \ldots  &   \ldots   & \ldots    & \ldots       & 5 \\
2047.65 & \phn 0.078 & \phn 0.053 &  \ldots   &  \ldots    &   \ldots  &   \ldots   & \ldots    & \ldots       & 5 \\
2055.63 &   \ldots   & \phn 0.075 &  \ldots   &  \ldots    &   \ldots  &   \ldots   & \ldots    & \ldots       & 5 \\
\hline
2052.66 & \phn 0.089 & \phn 0.068 & $-$0.050  & \phn 0.017 & \phn 0.018 & \phn 0.006 & \phn 0.000 & $-$ 0.028  & 6 \\
\enddata
\tablenotetext{a}{Values in Columns 2 through 9 are in magnitudes.  
The K-corrrections are subtracted from the data in Table 4. 
The S-corrections are added to the data in Table \ref{ubvri}. }
\tablenotetext{b}{Julian Date {\em minus} 2,450,000.}
\tablenotetext{c}{1 = CTIO 0.9-m, 2 = KAIT, 3 = Nickel 1-m, 4 = Yale-AURA-Lisbon-Ohio (YALO) 1-m,
5 = LCO 2.5-m, and 6 = CTIO 1.5-m telescopes.}
\end{deluxetable}

\begin{deluxetable}{cccc}
\tablewidth{0pt}
\tablecolumns{4}
\tablecaption{Near-Infrared Photometry of SN~2001ay\tablenotemark{a}\label{jh}}
\tablehead{
\colhead{JD\tablenotemark{b}}  &
\colhead{$J_s$ (mag)} &
\colhead{$H$ (mag)} &
\colhead{Telescope\tablenotemark{c}}
}
\startdata
2024.73     &    \ldots        &  17.045 (0.042) & 1 \\
2025.67     &  16.977 (0.032)  &  16.893 (0.044) & 1 \\
2026.70     &  17.050 (0.028)  &   \ldots        & 1 \\
2029.70     &  17.132 (0.022)  &  17.000 (0.035) & 1 \\
2031.68     &  17.136 (0.022)  &  16.965 (0.036) & 1 \\
2032.70     &  17.190 (0.028)  &  17.110 (0.027) & 1 \\
2045.66     &  17.291 (0.024)  &  16.873 (0.020) & 2 \\
2047.65     &  17.509 (0.049)  &   \ldots        & 1 \\
2049.66     &  17.396 (0.035)  &  16.853 (0.039) & 1 \\
2051.65     &  17.430 (0.034)  &  16.894 (0.038) & 1 \\
2053.67     &  17.337 (0.031)  &  \ldots         & 1 \\
2055.67     &  17.387 (0.031)  &  17.078 (0.056) & 1 \\
2056.67     &  17.420 (0.031)  &  16.998 (0.022) & 1 \\
2061.65     &    \ldots        &  16.905 (0.022) & 2 \\
2068.57     &  17.592 (0.033)  &  17.157 (0.028) & 2 \\
\enddata
\tablenotetext{a}{Magnitude uncertainties ($1\sigma$) are given
in parentheses.}
\tablenotetext{b}{Julian Date {\em minus} 2,450,000.}
\tablenotetext{c}{1 = LCO 1-m, 2 = LCO 2.5-m telescopes.}
\end{deluxetable}

\begin{deluxetable}{lccccccccccccl}
\tabletypesize{\scriptsize}
\rotate
\tablewidth{650pt}
\tablecaption{Spectroscopic Observations in the ``CfA Set''\label{sp1_table}}
\tablehead{\colhead{UT Date} &
\colhead{Julian Day} &
\colhead{Tel.} &
\colhead{Range}  &
\colhead{Res.} &
\colhead{Pos.A.} &
\colhead{Par.A.} &
\colhead{Airmass} & 
\colhead{Flux Std.} &
\colhead{Sky} &
\colhead{Seeing} &
\colhead{Slit} &
\colhead{Exp.} &
\colhead{Observer(s)} \\
\colhead{(2001)} &
\colhead{} &
\colhead{} &
\colhead{(\AA)} &
\colhead{(\AA)} &
\colhead{($^\circ$)} &
\colhead{($^\circ$)} &
\colhead{} &
\colhead{} &
\colhead{} &
\colhead{($^{\prime\prime}$)} &
\colhead{($^{\prime\prime}$)} &
\colhead{(s)} & 
\colhead{} }
\startdata
Apr 21.48 & 2452020.98 & FLWO &    3720--7540 & 7.0 &   71.00 &   70.02 & 1.48 &    Feige34 &            &    1--2 & 3 & 1200 &          Calkins \\
Apr 24.31 & 2452023.81 & FLWO &    3720--7540 & 7.0 &    0.00 &  $-$16.25 & 1.01 &    Feige34 &      clear &      2 & 3 & 1200 &          Berlind \\
Apr 26.33 & 2452025.83 & FLWO &    3720--7540 & 7.0 &   45.00 &   43.51 & 1.01 &    Feige34 & some clouds &    1--2 & 3 & 1200 &          Berlind \\
Apr 26.40 & 2452025.90 & MMTO &      3250--8900 & 8.0 & Par.A. &   71.15 & 1.12 &    Feige34/BD26 &            &   &   &  900 &       Matheson \\
Apr 26.41 & 2452025.91 & MMTO &      3250--8900 & 8.0 & Par.A. &   71.52 & 1.15 &    Feige34/BD26 &            &   &   &  900 &    Matheson \\
Apr 27.36 & 2452026.86 & FLWO &    3720--7540 & 7.0 &   64.00 &   65.79 & 1.04 &       HZ44 &            &    1--2 & 3 & 1200 &          Calkins \\
Apr 30.36 & 2452029.86 & FLWO &    3720--7540 & 7.0 &   70.00 &   67.86 & 1.05 &    Feige34 &      clear &    1--2 & 3 & 1200 &          Berlind \\
May 02.29 & 2452031.79 & FLWO &    3720--7540 & 7.0 &   $-$9.00 &  $-$19.75 & 1.01 &       HZ44 &      clear &    1--2 & 3 & 1200 &          Berlind \\
May 16.26 & 2452045.76 & FLWO &    3720--7540 & 7.0 &  $-$33.00 &  $-$13.29 & 1.01 &       HZ44 &      clear &    1--2 & 3 & 1200 &          Calkins \\
May 23.33 & 2452052.83 & FLWO &    3720--7540 & 7.0 &   69.00 &   71.42 & 1.14 &    Feige34 &     cirrus &      2 & 3 & 1200 &            Rines \\
May 24.32 & 2452053.82 & MMTO &      3220--8900 & 8.0 & at Par.A. &   70.92 & 1.11 & BD28/BD26 & clear      &         &   & 1200 & Challis, Phelps    \\
May 24.33 & 2452053.83 & MMTO &     4950--10000 & 8.0 & Par.A. &   71.51 & 1.16 &       BD26 & clear      &        &   & 1200 & Challis, Phelps    \\
May 25.32 & 2452054.82 & MMTO &      3220--8900 & 8.0 & Par.A. &   71.36 & 1.14 & Feige34/BD26 & clear      &      &   & 1800 & Challis \\
May 25.35 & 2452054.85 & MMTO &     4950--10000 & 8.0 & Par.A. &   71.52 & 1.23 &       BD26 & clear      &        &   & 1800 & Challis \\
May 25.37 & 2452054.87 & FLWO &      3720--7500 & 7.0 &   69.00 &   70.63 & 1.39 &    Feige34 &     cirrus &      2 & 3 & 1200 &            Rines \\
May 30.36 & 2452059.86 & FLWO &    3720--7540 & 7.0 &   71.00 &   70.51 & 1.40 &       HZ44 &      clear &    1--2 & 3 & 1200 &          Calkins \\
May 30.38 & 2452059.88 & FLWO &    3720--7540 & 7.0 &   71.00 &   69.68 & 1.53 &       HZ44 &     cirrus &    1--2 & 3 & 1200 &          Calkins \\
June 18.22 & 2452078.72 & FLWO &    3720--7540 & 7.0 &   69.00 &   66.99 & 1.05 &       HZ44 &   clouds   &    1--2 & 3 & 1200 &          Berlind \\
June 18.29 & 2452078.79 & MMTO &      3200--8900 & 8.0 & Par.A. &   71.24 & 1.29 &  BD28/BD26 & some clouds &      &   &  900 & Matheson           \\
June 18.30 & 2452078.80 & MMTO &      3200--8900 & 8.0 & Par.A. &   70.84 & 1.35 &  BD28/BD26 & some clouds &      &   &  900 & Matheson           \\
\enddata
\end{deluxetable}

\begin{deluxetable}{cclcc}
\tablewidth{0pt}
\tabletypesize{\scriptsize}
\tablecolumns{5}
\tablecaption{Other Spectroscopic Observations\label{sp2_table}}
\tablehead{
\colhead{UT Date (2001)} &
\colhead{Phase (d)\tablenotemark{a}} &
\colhead{Telescope} &
\colhead{Wavelength Range (\AA)} &
\colhead{Resolution (\AA)}
}
\startdata

  Apr 22 & \phn$-$1 & Keck II 10-m   &     3922--10000           & 1.25 \\
  Apr 29 &  \phn +6 & KPNO 2.1-m     &     5350--[9500]          & 4.8 \\
  Apr 30 &  \phn +7 & Lick 3-m       &     3260--10600           & 5.5/10.5 \\
  May 01 &  \phn +8 & LCO 2.5-m      &     3500--9199            & 7.0 \\
  May 02 &  \phn +9 & {\it HST} 2.4-m &  2877--5348            & 3.55 to 4.10 \\
  May 09 &      +16 & {\it HST} 2.4-m &  2930--5695            & 3.55 to 4.10 \\ 
  May 10 &      +17 & BAO 2.16-m     &     4100--9000            & 9.6 \\
  May 16 &      +23 & Lick 3-m       &     3300--10450           & 5.5/10.5 \\
  May 17 &      +24 & LCO 2.5-m      &     3500--9350            & 7.0 \\
  May 20 &      +27 & BAO 2.16-m     &     3450--9000            & 9.6 \\
  May 23 &      +30 & KPNO 4.0-m     & 3500--5493/5300--11600     & 7.8/13.8 \\
  May 25 &      +32 & BAO 2.16-m     &     3700--9300            & 9.6 \\
  May 26 &      +33 & LCO 2.5-m      &     3800--9234            & 7.0 \\
  May 30 &      +37 & BAO 2.16-m     &     3970--8200            & 4.8 \\
\enddata
\tablenotetext{a} {Number of observer frame days since time of $B$-band
maximum, 2001 April 23.}
\end{deluxetable}

\begin{deluxetable}{lccc}
\tablewidth{0pt}
\tablecolumns{5}
\tablecaption{Comparison of Three Objects\label{three}}
\tablehead{
\colhead{Parameter} &
\colhead{SN~2001ay} &
\colhead{SN~2005eq} &
\colhead{SN~2009dc} 
}
\startdata
\dmm\       &    0.68 (0.05)       &    0.72--0.78     &    0.72 (0.03)  \\
$M_V$(max) &    $-$19.17 (0.10)   &    $-$19.40 (0.08)  &   $-$20.16 (0.10) \\
Si~II velocity\tablenotemark{a} (\kms)  &   $-$14,340 [$-$1.5~d]  & $-$10,200 [$-$5~d] & $-$8600 [$-$7 d]  \\
C~II line(s) &    weak              &  not detectable    & very strong \\
$H$-band light curve   &   flat   &  peaked early & increasing \\
\enddata
\tablenotetext{a}{For the $\lambda_0$ = 6355~\AA\ line.  
The values in brackets correspond to the numbers of rest-frame 
days after the time of $B$ maximum when the line was measured
at this velocity. For SN~2009dc the velocity is from \citet{Sil_etal10}.}
\end{deluxetable}

\clearpage

\figcaption[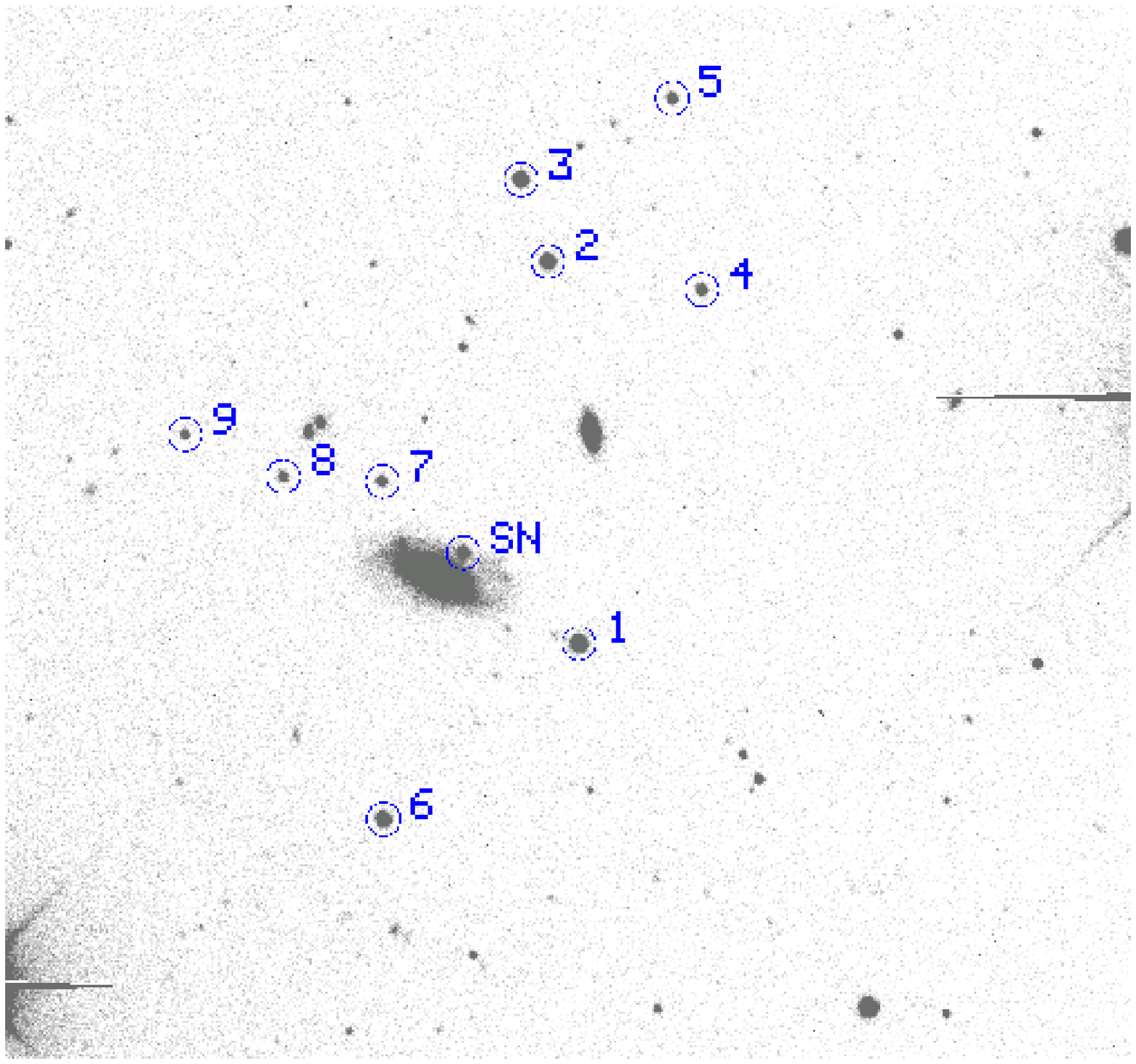] {Finder chart for IC 4423, SN~2001ay,
  and some field stars in our Galaxy.  
\label{finder}
}

\figcaption[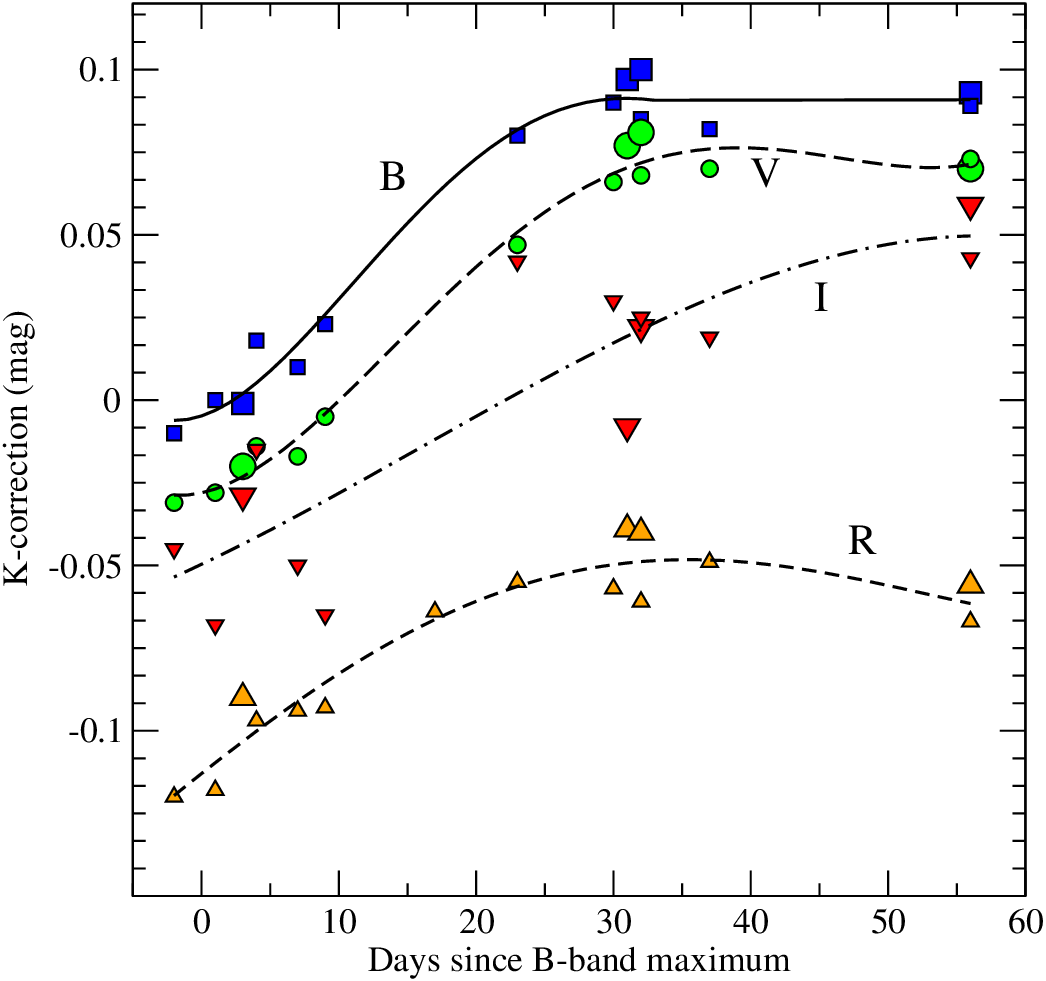]{K-corrections for $BVRI$ photometry of
  SN~2001ay. 
\label{kcorr}
}

\figcaption[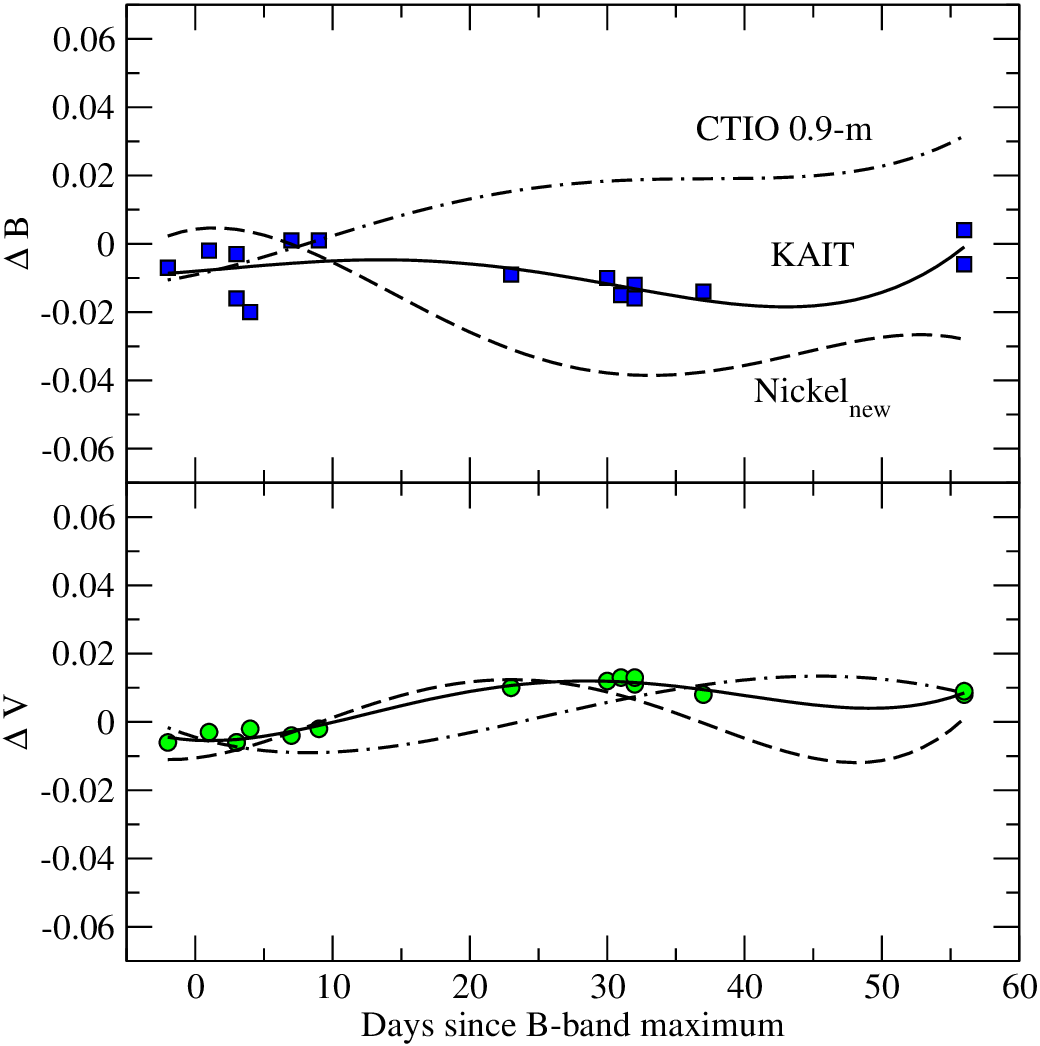] {S-corrections for $B$- and $V$-band
  photometry of SN~2001ay.
\label{scorr_bv}
}

\figcaption[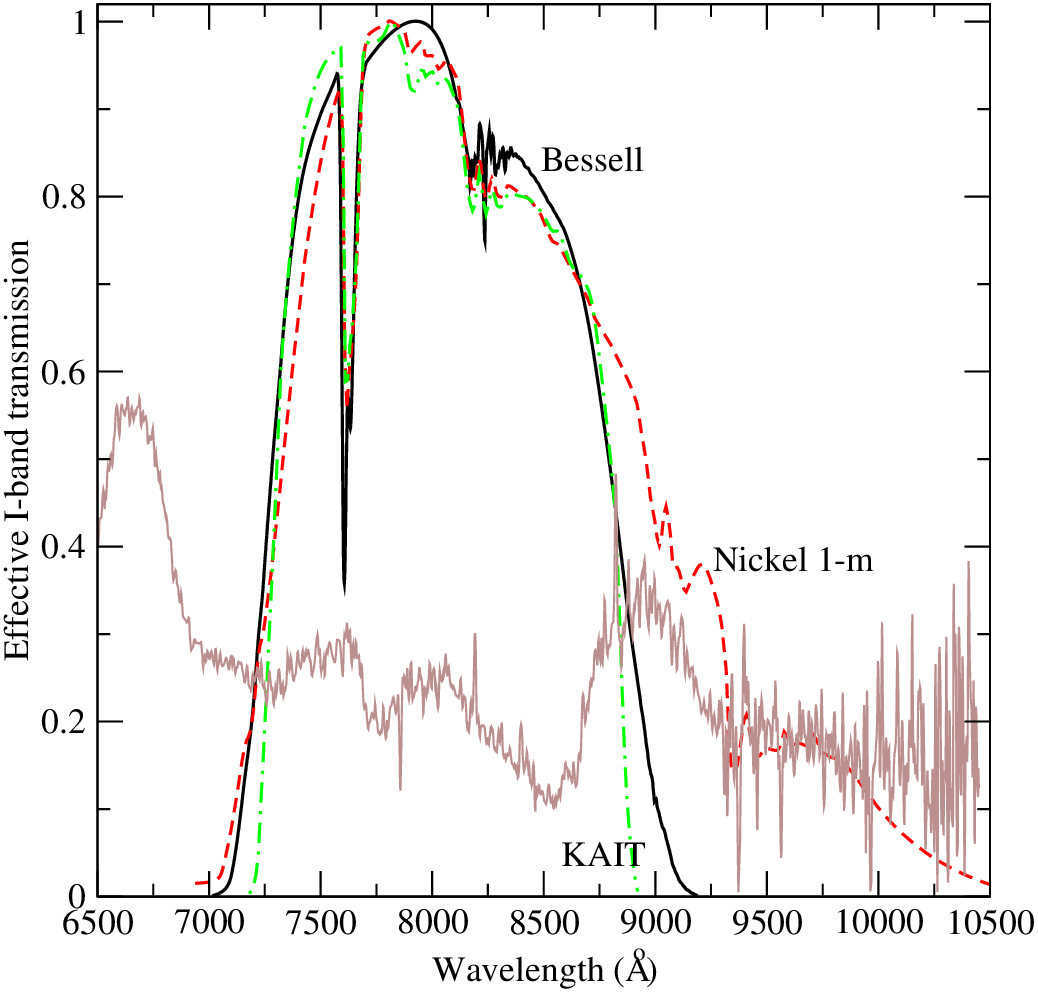]{Effective $I$-band transmission functions for
  the \citet{Bes90} filter, the KAIT filter, and the Nickel 1-m filter.
\label{i_filter}
}

\figcaption[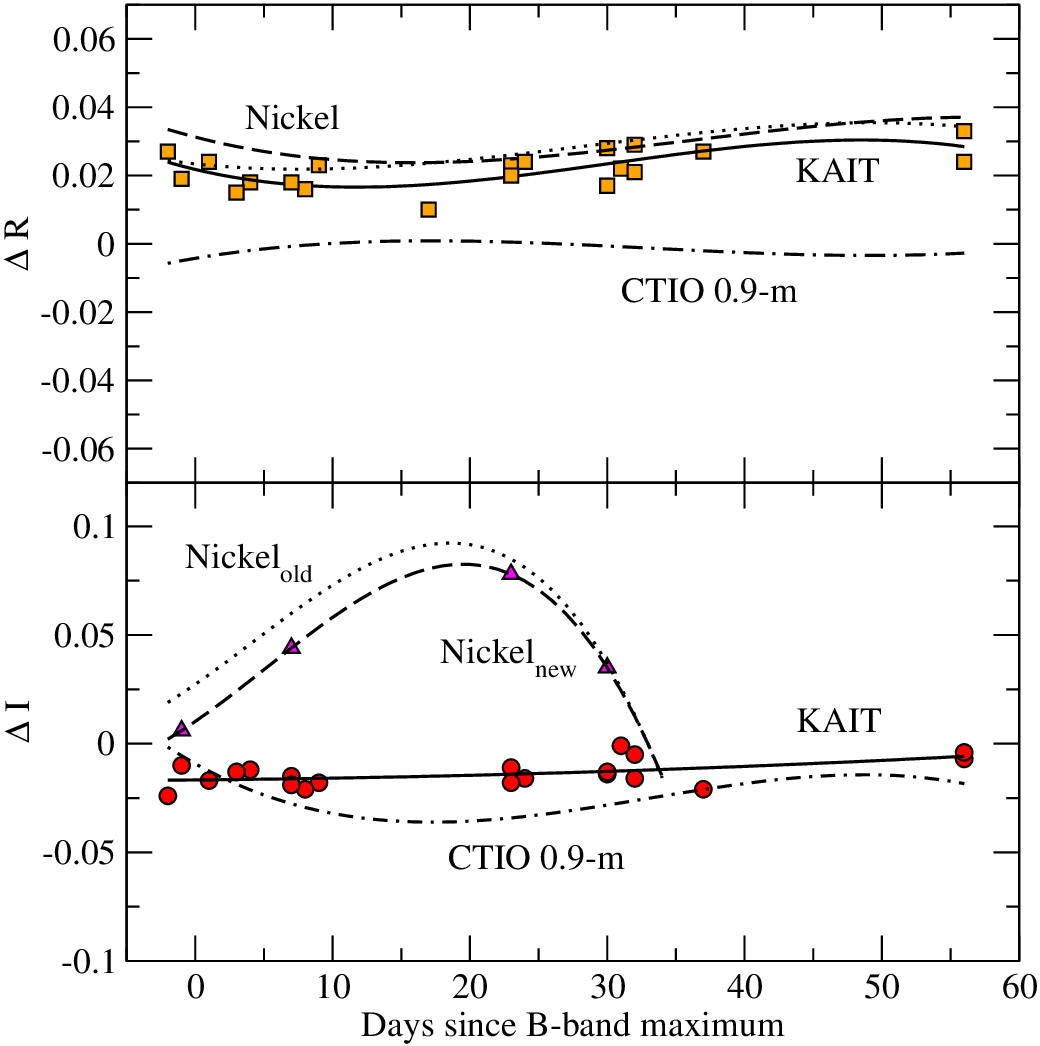] {S-corrections for $R$- and $I$-band
  photometry of SN~2001ay.  
\label{scorr_ri}
}

\figcaption[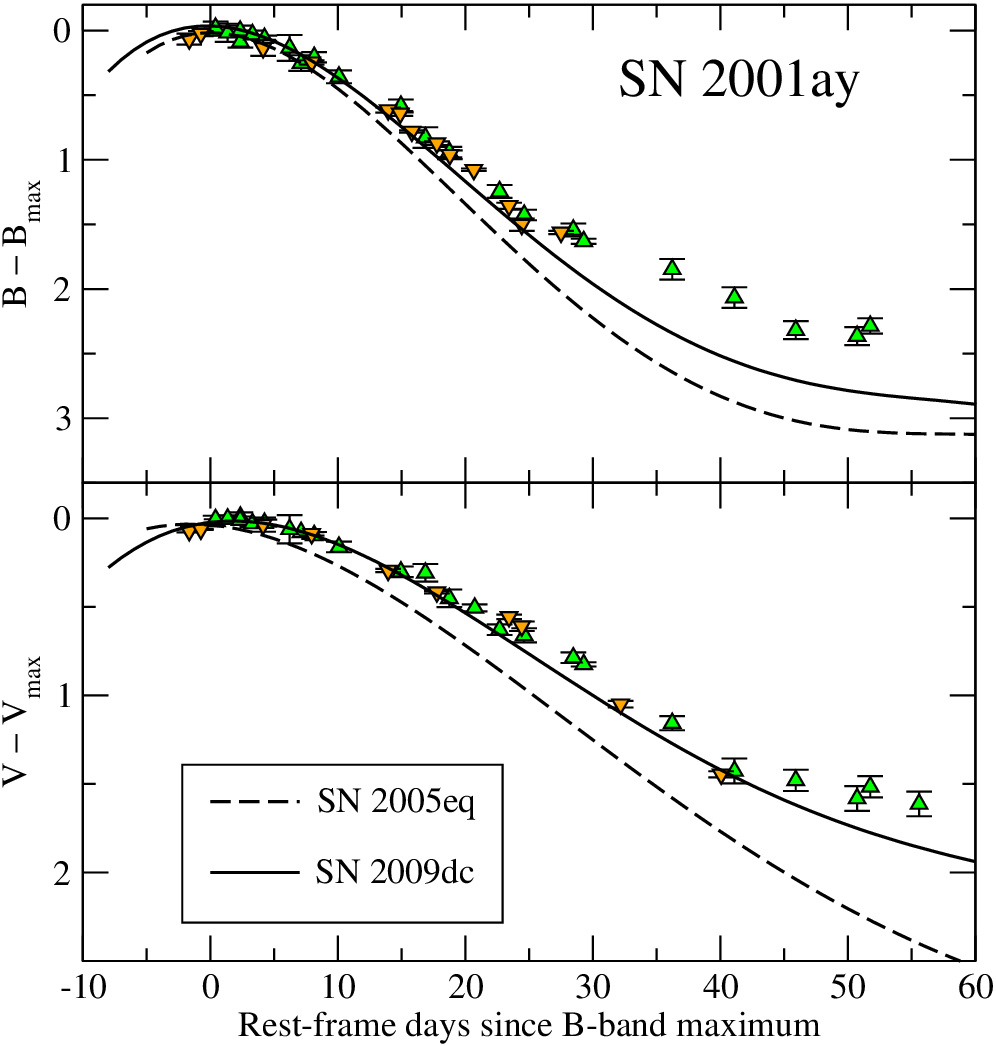] {Comparison of the $B$- and $V$-band light
  curves of SNe~2001ay, 2005eq \citep{Con_etal10}, and 2009dc
  \citep{Sil_etal10}.  
\label{bv_comp}
}

\figcaption[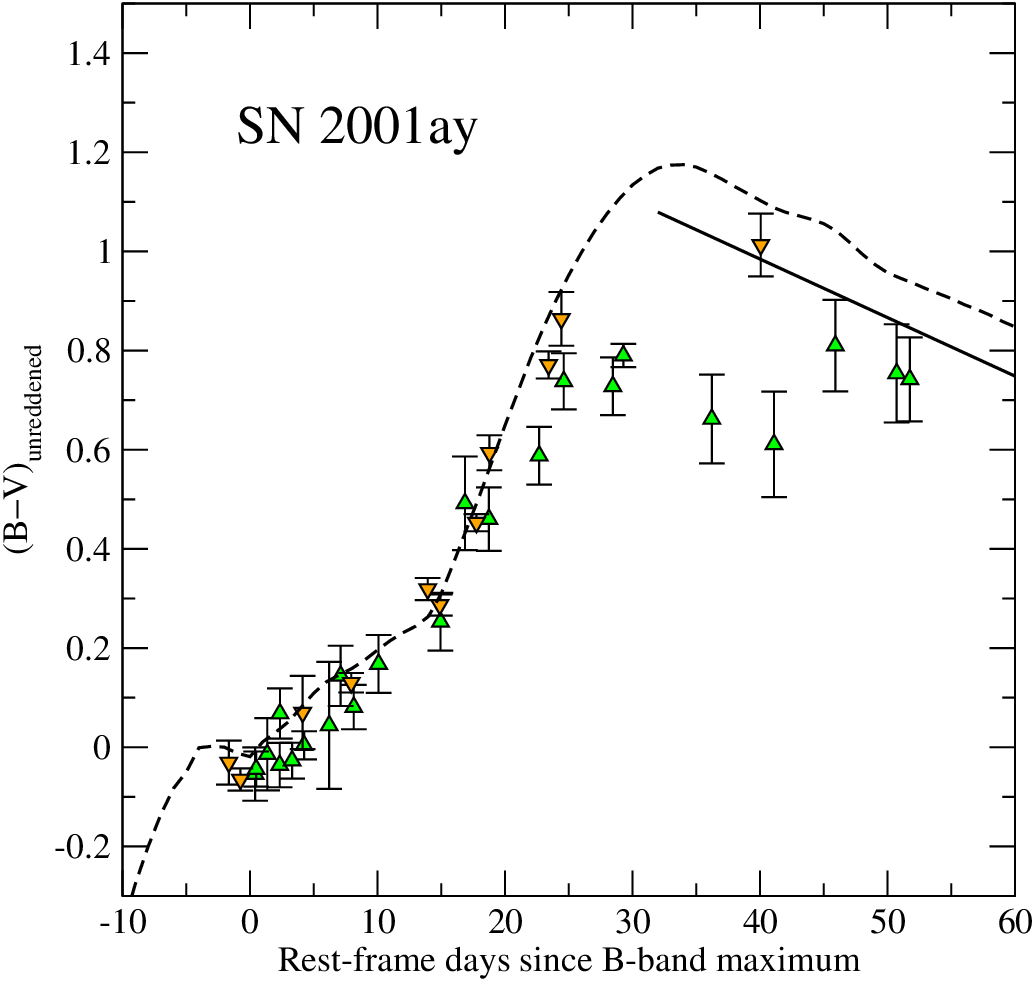] {Unreddened $B-V$ colors of SN~2001ay vs.
  time since $B$-band maximum.  
\label{01ay_bv}
}

\figcaption[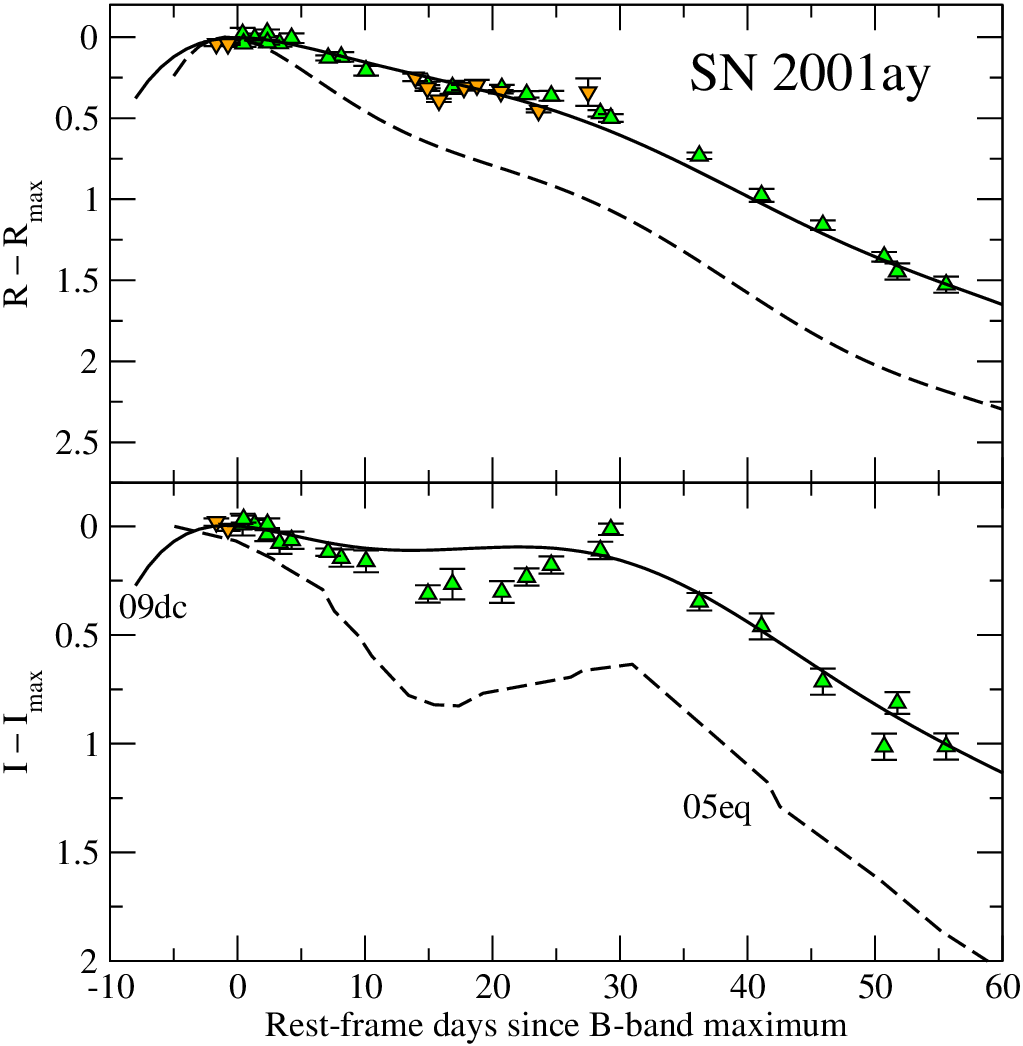] {Similar to Figure \ref{bv_comp}, except for
  the $R$ and $I$ bands.  
\label{ri_comp}
}

\figcaption[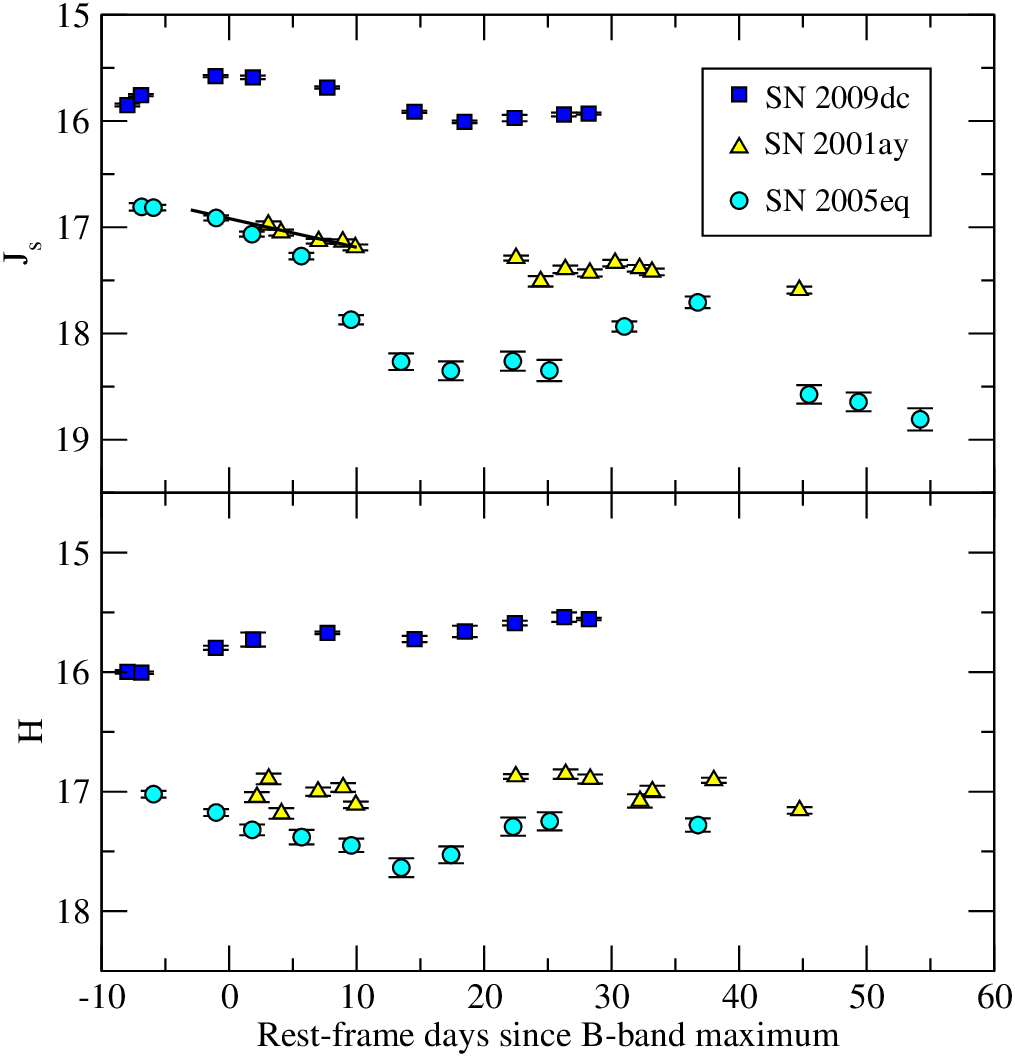] {Near-IR photometry of SNe~2009dc
  \citep{Str_etal11}, 2001ay, and 2005eq \citep{Con_etal10}.  
\label{jh_comp}
}

\figcaption[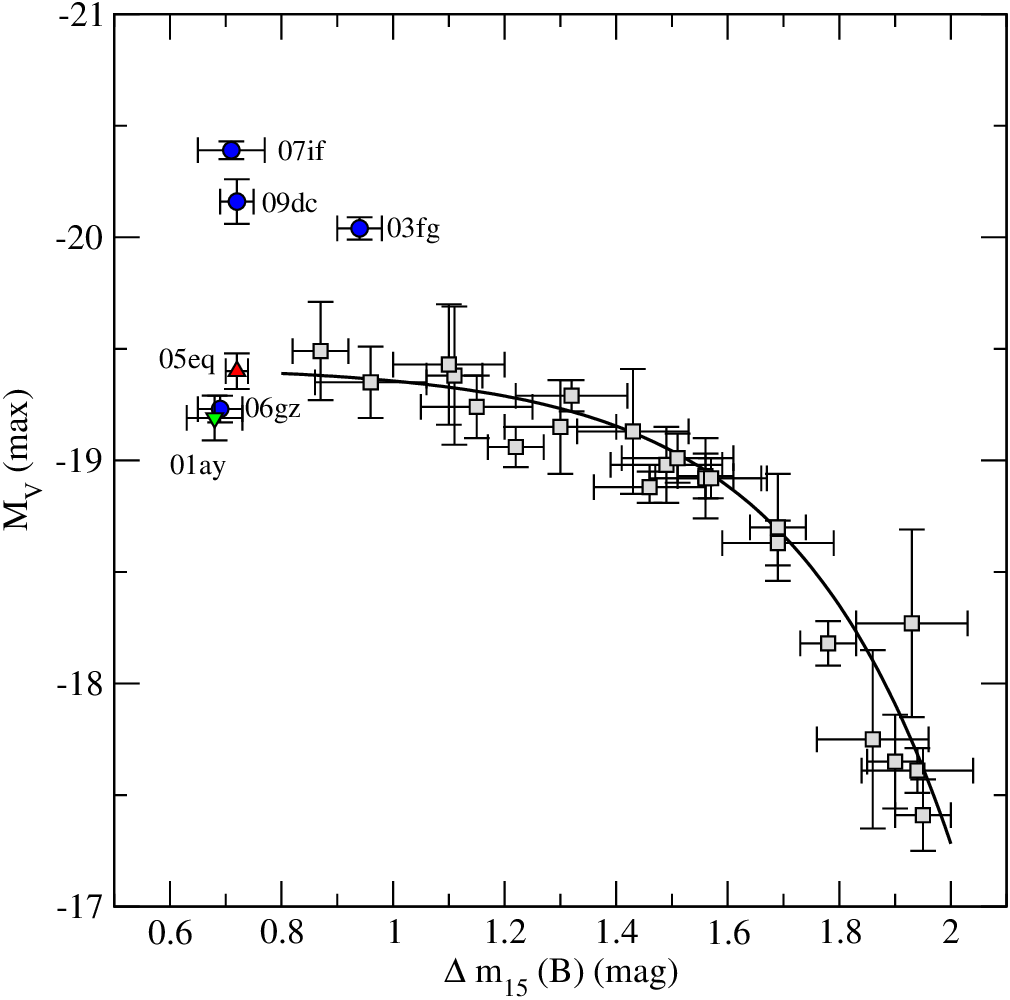] {Absolute $V$-band magnitudes 
  at maximum light of six
  very slowly declining Type Ia SNe vs. the decline-rate parameter
  \dmm.  The decline-rate relation of \citet{Gar_etal04} is also
  shown.
\label{slow_decliners}
}

\figcaption[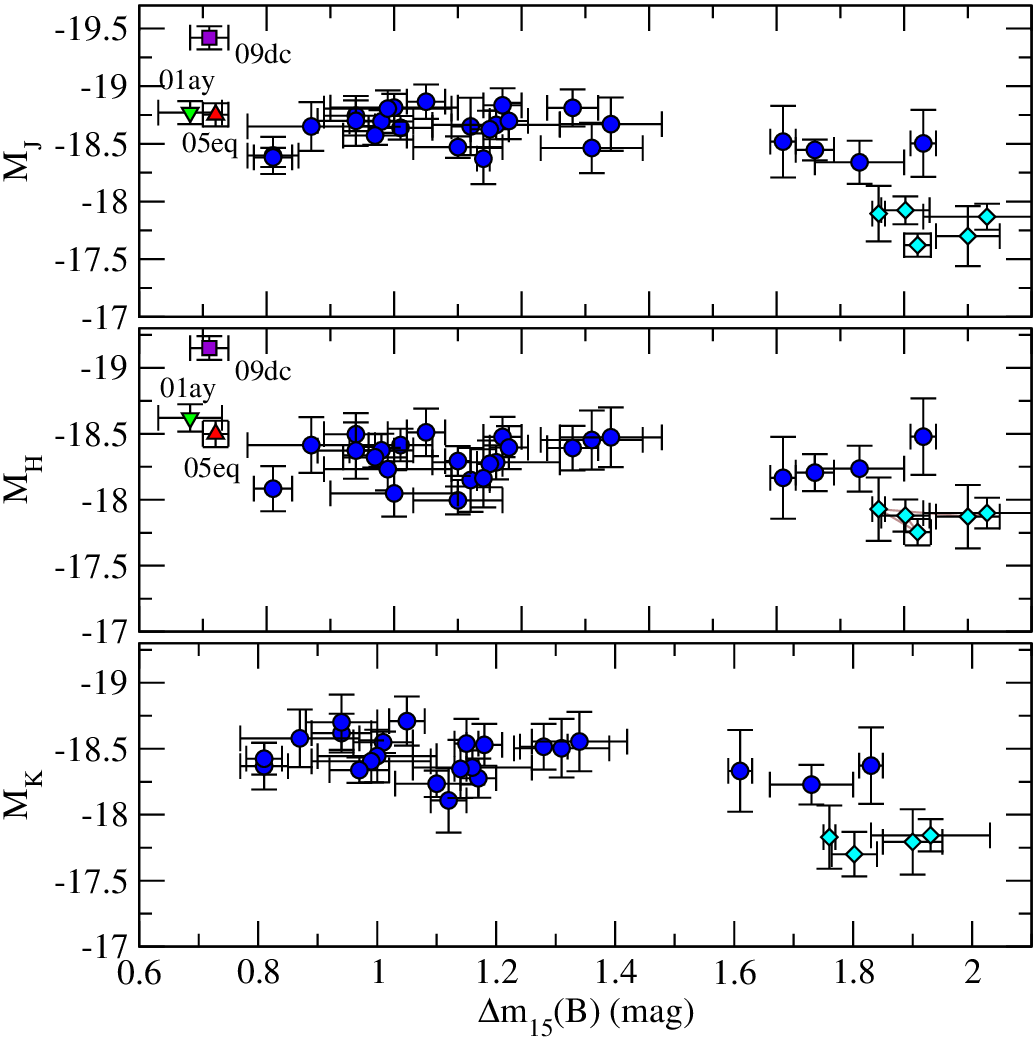] {Near-IR absolute magnitudes at maximum
  light of Type Ia SNe.  
\label{absmag_ir}
}

\figcaption[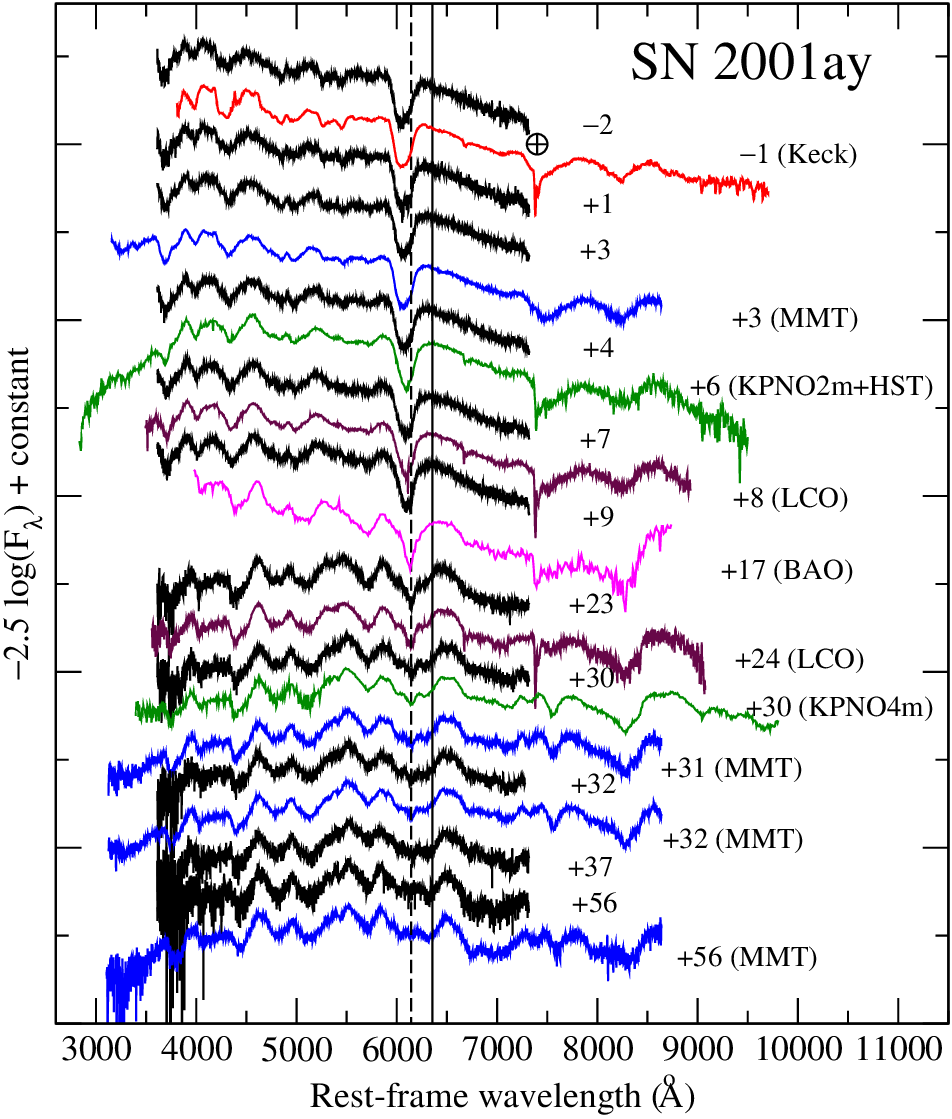] {Temporal sequence of spectra of
  SN~2001ay. 
\label{spectra}
}

\figcaption[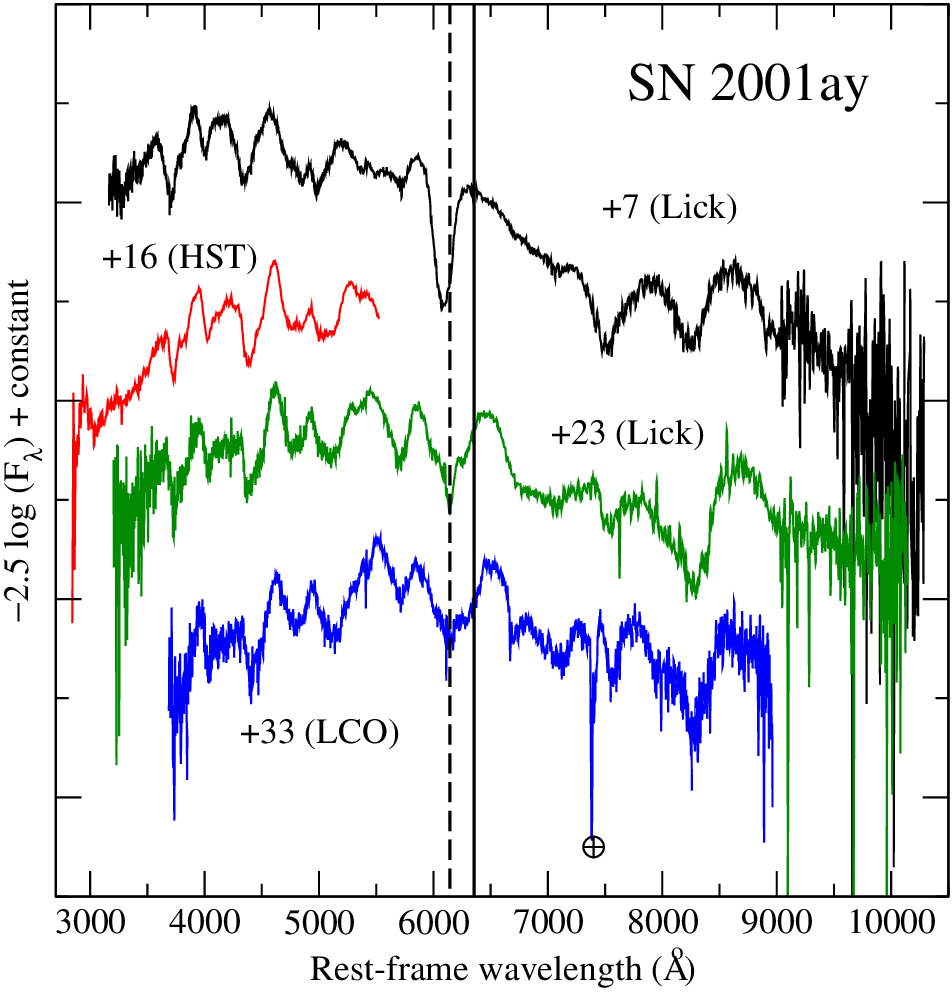] {Additional spectra of SN~2001ay
  obtained with the Lick Observatory 3-m Shane telescope, {\it HST},
  and the Las Campanas Observatory 2.5-m du Pont telescope.  
\label{additional_spectra}
}

\figcaption[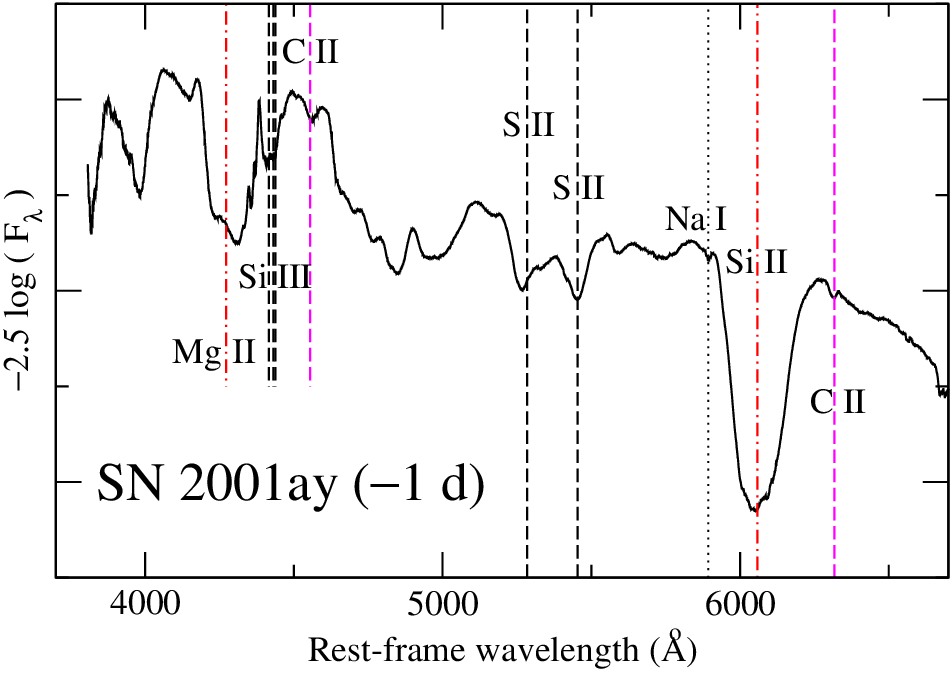] {A portion of our highest signal-to-noise ratio
  spectrum of SN~2001ay, taken with the Keck ESI on 2001 April 22.
\label{keck}
}

\figcaption[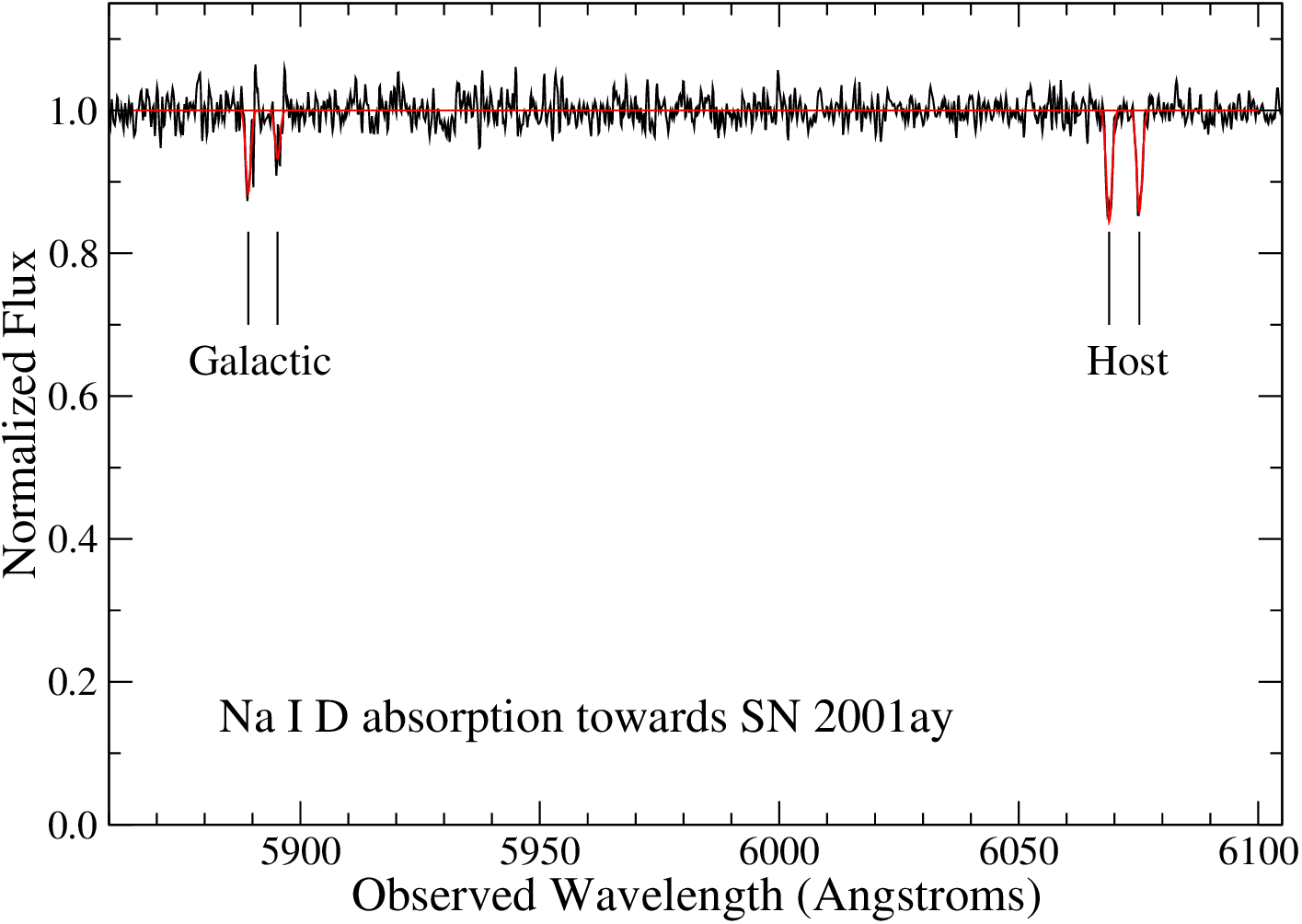] {Profile of the Na~I~D lines in our Keck
  ESI spectrum.  The Milky Way and host-galaxy components are clearly
  present.
\label{sodium}
}

\figcaption[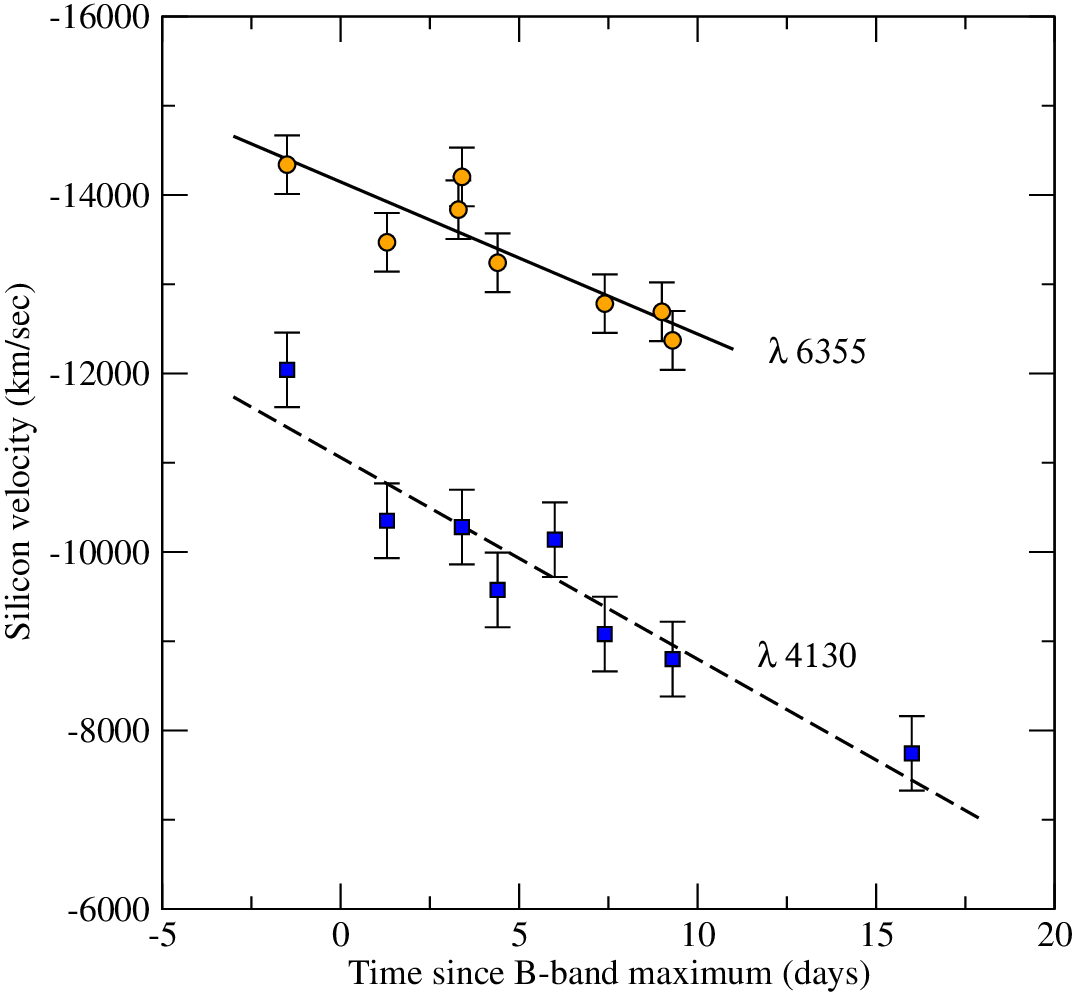] {Blueshifted velocity of two Si~II lines in
  the spectra of SN~2001ay around maximum light.  
\label{silicon}
}

\figcaption[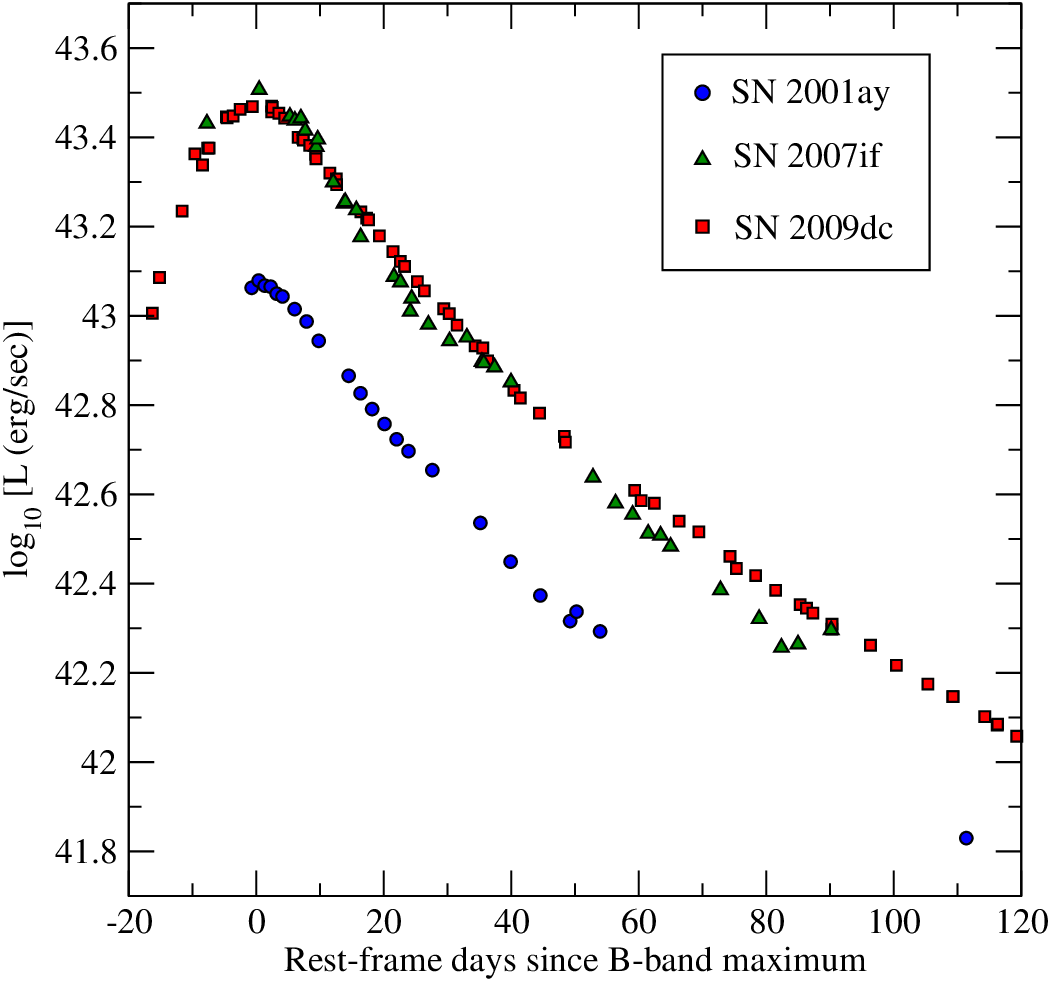] {Bolometric luminosities ($L$) of SNe~2001ay,
  2007if \citep{Sca_etal10}, and 2009dc \citep{Tau_etal11},
  measured in erg s$^{-1}$.  
\label{bolom_lc}
}

\figcaption[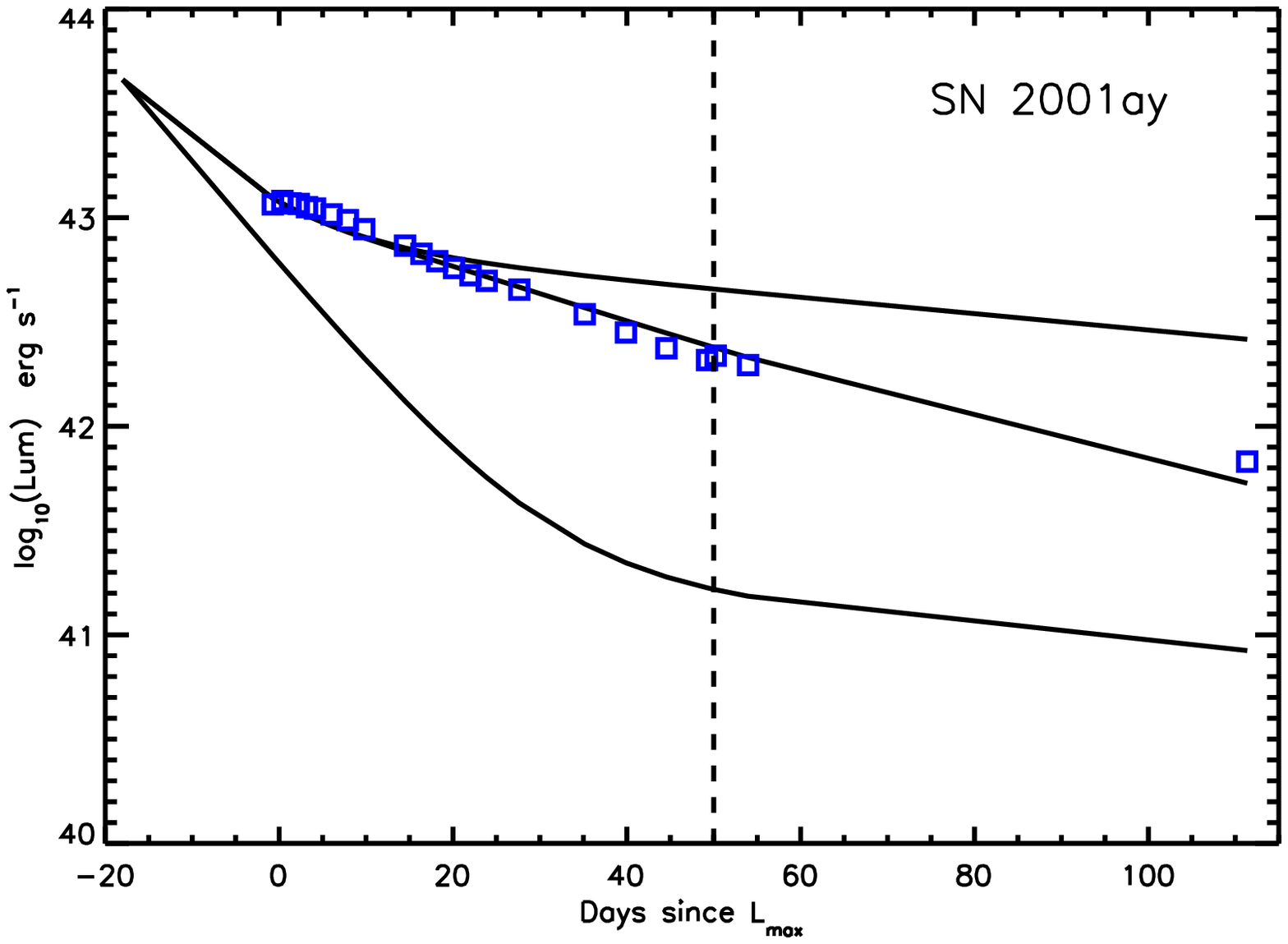] {The best-fit radioactive decay energy deposition
function (middle solid line) to the UVOIR light curve (blue squares) of
SN~2001ay.  
\label{edep}
}

\figcaption[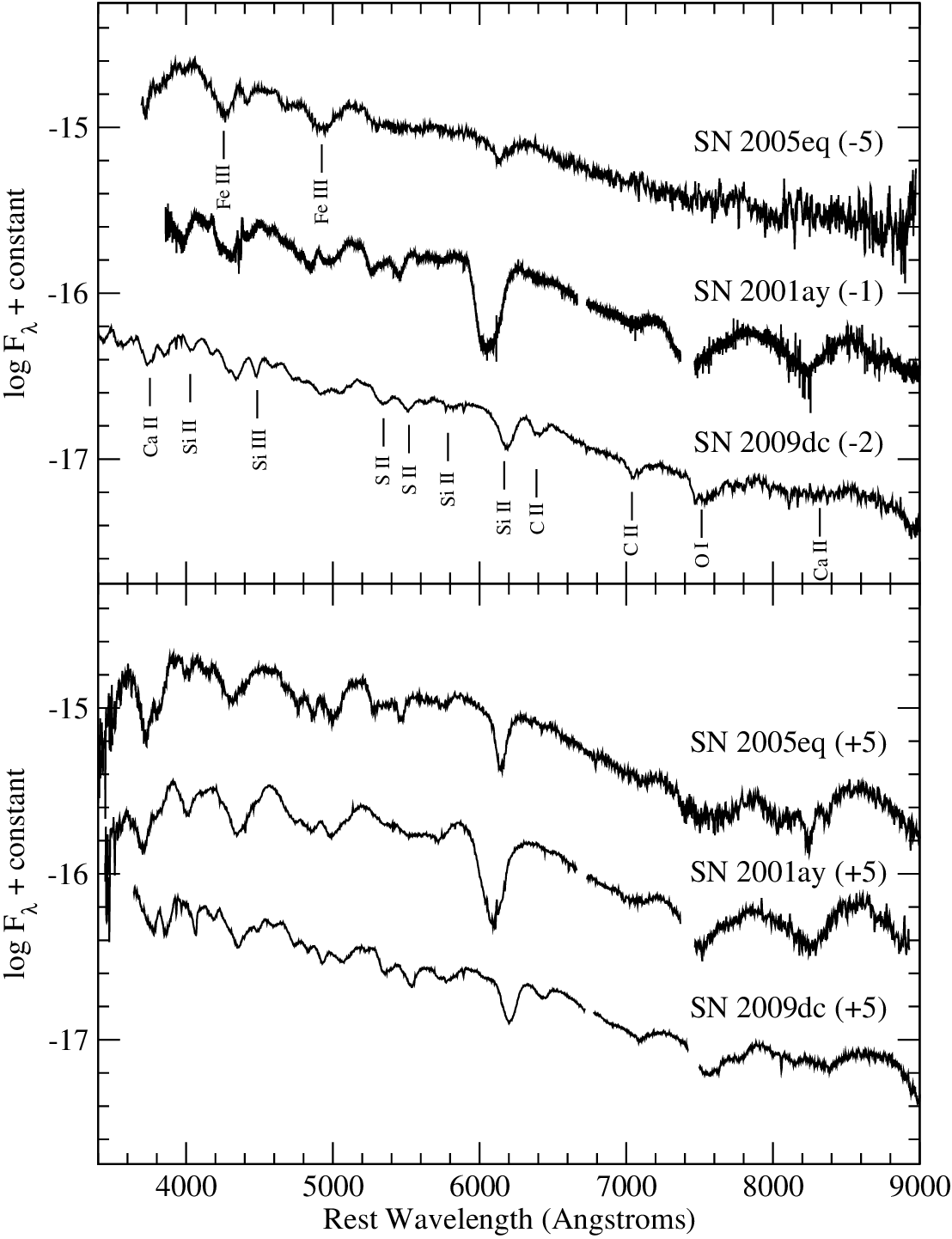] {Comparison of the optical spectra of
  SNe~2001ay, 2005eq, and 2009dc.
\label{3spectra}
}

\figcaption[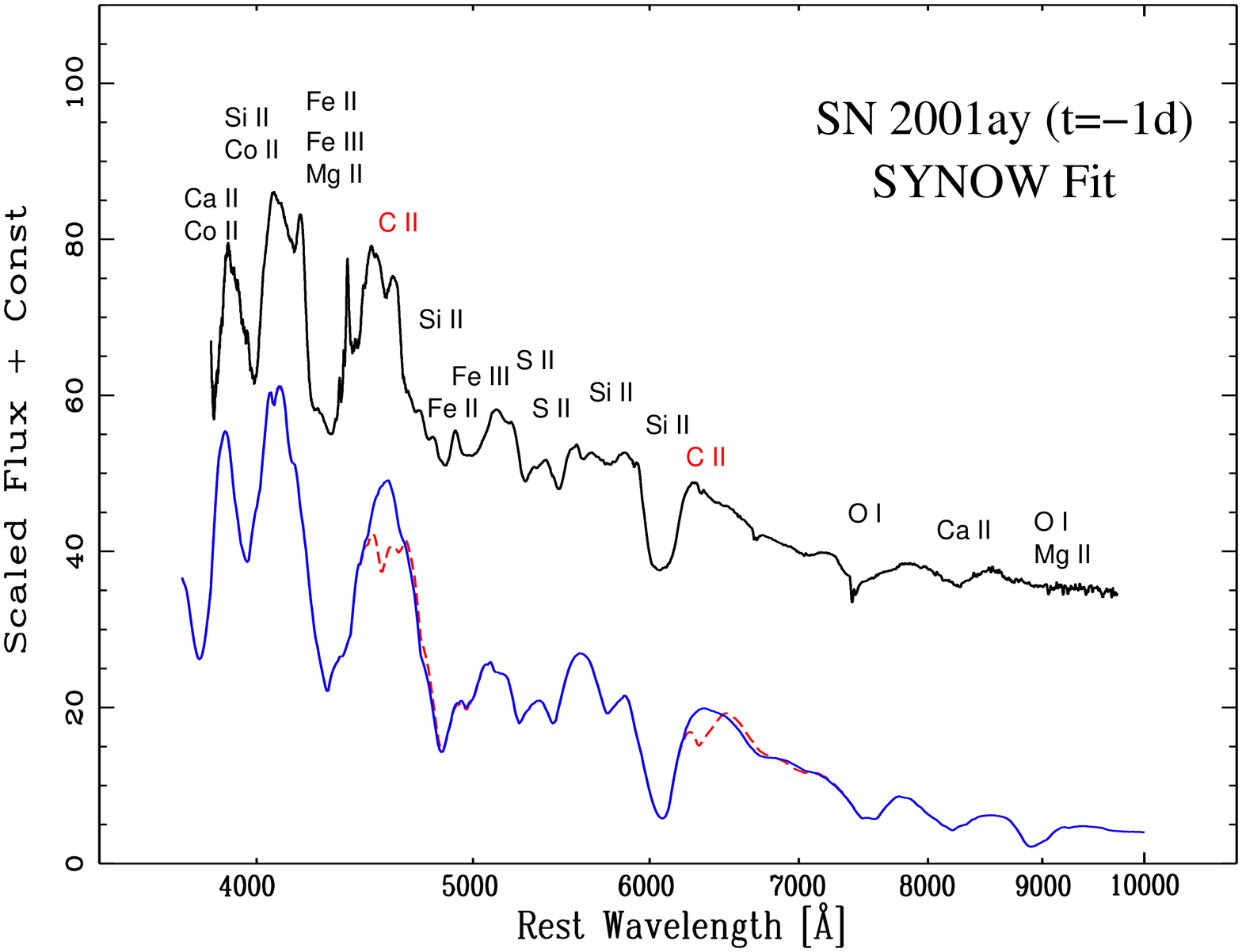] {SYNOW modeling of the optical spectrum of
  SN~2001ay at 1~d before $B$ maximum. Upper curve = data; lower
  curve = SYNOW model.  The dashed lines indicate the addition of
  C~II to the model.  
\label{synow1}
}

\figcaption[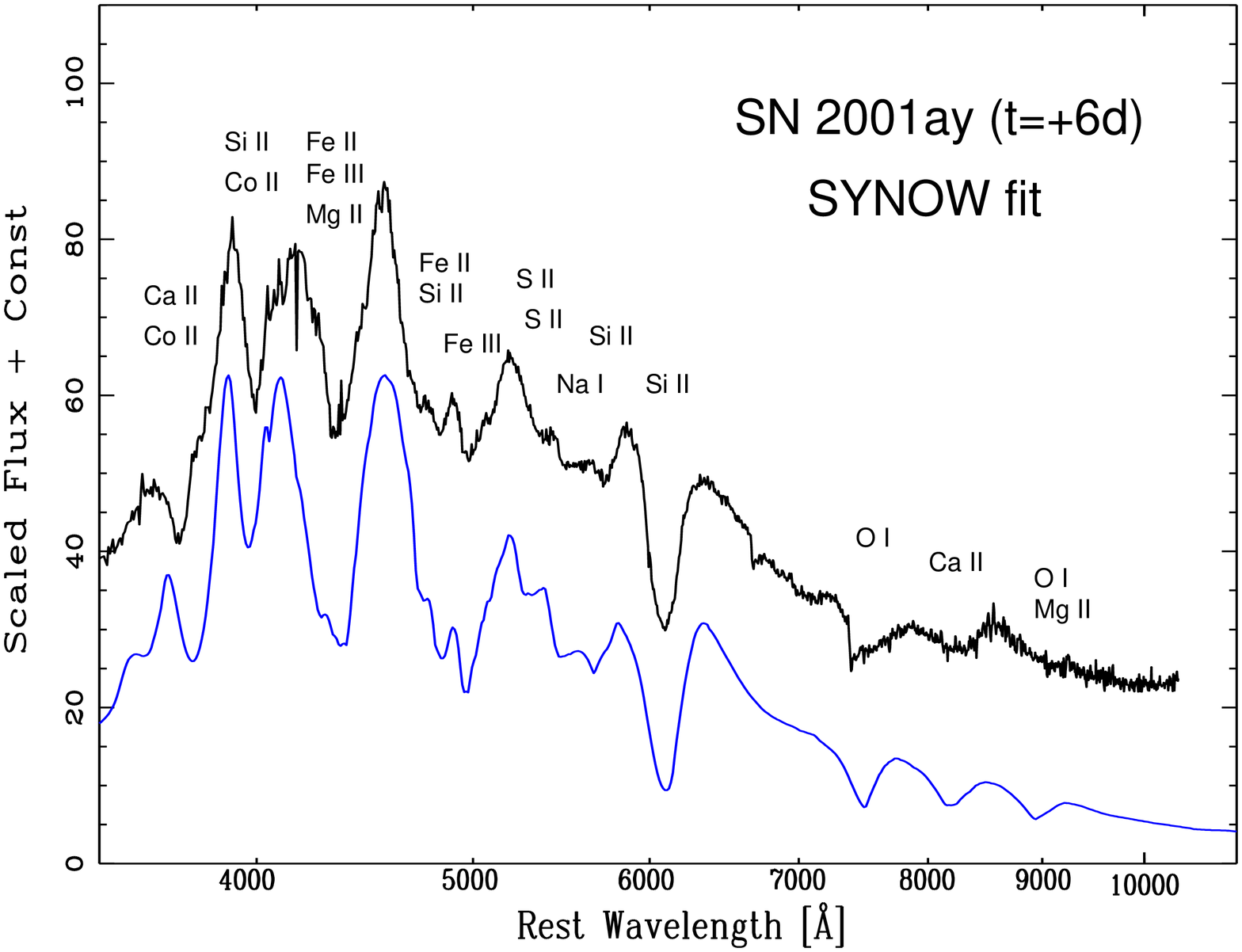] {SYNOW modeling of the optical spectrum of
  SN~2001ay 6~d after $B$ maximum.  Upper curve = data; lower
  curve = SYNOW model.
\label{synow2}
}

\figcaption[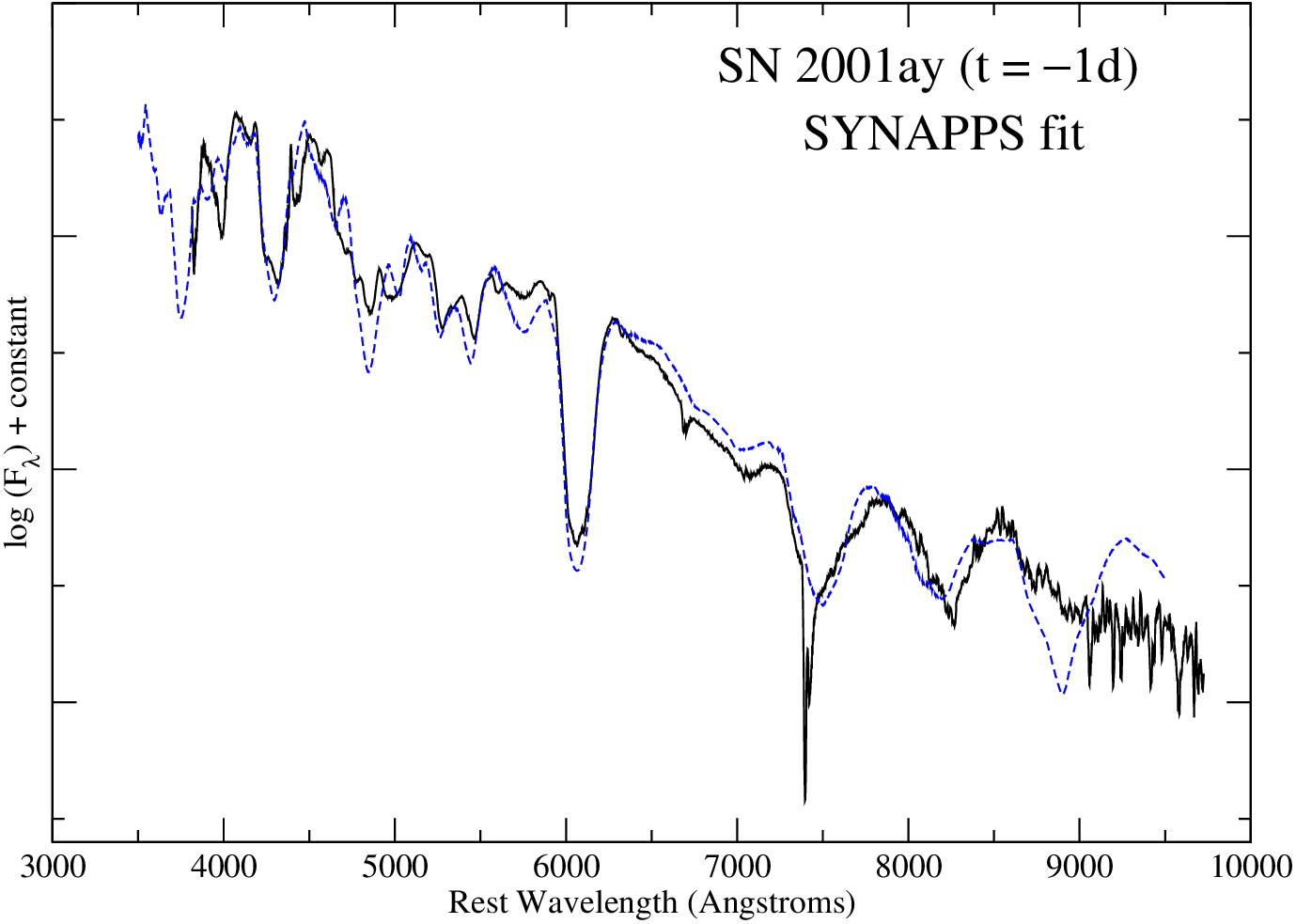] {SYNAPPS modeling of the optical spectrum of
  SN~2001ay 1~d before $B$ maximum.  The blue dashed line (model
  spectrum) can be shifted arbitrarily along the vertical axis to match
  the actual spectrum (shown in black).
\label{synapps}
}


\begin{figure}
\plotone{sn2001ay_finder.ps}
{\center Krisciunas {\it et al.} Figure \ref{finder}.
Finder chart for IC 4423, SN~2001ay,
  and some field stars in our Galaxy.  This $7.4' \times 7.4'$
  image is a 200~s $V$-band exposure obtained on 2001 May 23 with
  the CTIO 1.5-m telescope.  North is up and east is to the left.  The
  SN is located $10.3''$ W and $9.3''$ N of the galaxy
  core.}
\end{figure}

\begin{figure}
\plotone{kcorr.eps}
{\center Krisciunas {\it et al.} Figure \ref{kcorr}.
K-corrections for $BVRI$ photometry of
  SN~2001ay. These account for the shifting of the SN spectrum to
  longer wavelengths owing to the redshift of the host galaxy.  These
  values are to be {\em subtracted} from the photometry.  Values
  derived from the high signal-to-noise ratio MMT spectra are plotted
  as larger symbols.}
\end{figure}

\begin{figure}
\plotone{scorr_bv.eps}
{\center Krisciunas {\it et al.} Figure \ref{scorr_bv}.
S-corrections for $B$- and $V$-band
  photometry of SN~2001ay, based on spectra from the FLWO 1.5-m and
  the MMT.  The individual points are shown only for the KAIT
  corrections.  The corrections for the older Nickel 1-m chip are  
  essentially the same as those shown for the newer Nickel 1-m chip.
  S-corrections are {\em added} to the photometry to correct it to the
  filter system of \citet{Bes90}.}
\end{figure}

\begin{figure}
\plotone{Iband.eps}
{\center Krisciunas {\it et al.} Figure \ref{i_filter}.
Effective $I$-band transmission functions for
  the \citet{Bes90} filter, the KAIT filter, and the Nickel 1-m filter
  used with the newer, higher quantum efficiency chip.  Each filter
  profile has been normalized to 1.00 at maximum throughput.  We also
  include the $t$ = +23~d spectrum from Figure
  \ref{additional_spectra} (multiplied by $10^{15}$); it shows
  significant flux past 9000~\AA, which is measured by the Lick 1-m
  system but excluded in the KAIT measurements.}
\end{figure}

\begin{figure}
\plotone{scorr_ri.eps}
{\center Krisciunas {\it et al.} Figure \ref{scorr_ri}.
S-corrections for $R$- and $I$-band
  photometry of SN~2001ay.  Corrections for the older Nickel 1-m chip
  are shown as a dotted line; corrections for the new Nickel 1-m chip
  are shown as a dashed line.  For the $I$-band corrections with the
  Nickel 1-m filter and both CCD chips used in the camera, we can only
  use spectra that extend to sufficiently long wavelengths (our Keck
  spectrum, two spectra taken with the Lick 3-m, and one with the KPNO
  4-m).  For the Nickel 1-m photometry to match the KAIT photometry,
  we would need the S-corrections to be a factor of 3 larger than the
  derived values. S-corrections are {\em added} to the photometry to
  correct it to the filter system of \citet{Bes90}.}
\end{figure}

\begin{figure}
\plotone{bv_comp.eps}
{\center Krisciunas {\it et al.} Figure \ref{bv_comp}.
Comparison of the $B$- and $V$-band light
  curves of SNe~2001ay, 2005eq \citep{Con_etal10}, and 2009dc
  \citep{Sil_etal10}.  The SN~2001ay data are K-corrected and
  S-corrected, and the CTIO 0.9-m and Nickel 1-m aperture photometry
  includes additional corrections discussed in the text.  For
  SN~2001ay the upward pointing triangles represent data which result
  from PSF photometry using image-subtraction templates.  The downward
  pointing triangles are SN~2001ay data based on aperture photometry.}
\end{figure}

\begin{figure}
\plotone{01ay_bv.eps}
{\center Krisciunas {\it et al.} Figure \ref{01ay_bv}.
Unreddened $B-V$ colors of SN~2001ay vs.
  time since $B$-band maximum.  Upward pointing triangles are data
  derived using image-subtraction templates.  Downward pointing
  triangles represent data derived from aperture magnitudes without
  template subtractions.  The dashed line is the locus from
  \citet{Pri_etal06b} for their most slowly declining object (with
  \dmm\ = 0.83 mag).  The solid line is the commonly used
  Lira-Phillips locus \citep{Lir95, Phi_etal99}.}
\end{figure}

\begin{figure}
\plotone{ri_comp.eps}
{\center Krisciunas {\it et al.} Figure \ref{ri_comp}.
Similar to Figure \ref{bv_comp}, except for
  the $R$ and $I$ bands.  In the case of SN~2005eq the data are taken
  in Sloan $r$ and $i$ filters, not in Kron-Cousins $R$ and $I$.  The
  Nickel 1-m $I$-band data are not shown.}
\end{figure}

\begin{figure}
\plotone{jh_comp.eps}
{\center Krisciunas {\it et al.} Figure \ref{jh_comp}.
Near-IR photometry of SNe~2009dc
  \citep{Str_etal11}, 2001ay, and 2005eq \citep{Con_etal10}.  The
  solid line is an educated guess that SN~2001ay may have been
  $\sim 0.12$ mag brighter in $J_s$ at maximum light than our
  earliest $J_s$ datum.}
\end{figure}

\begin{figure}
\plotone{slow_decliners.eps}
{\center Krisciunas {\it et al.} Figure \ref{slow_decliners}.
Absolute $V$-band magnitudes
  at maximum light of six
  very slowly declining Type Ia SNe vs. the decline-rate parameter
  \dmm.  The decline-rate relation of \citet{Gar_etal04} is also
  shown.}
\end{figure}

\begin{figure}
\plotone{absmag17.eps}
{\center Krisciunas {\it et al.} Figure \ref{absmag_ir}.
Near-IR absolute magnitudes at maximum
  light of Type Ia SNe.  While SNe 2001ay and 2005eq are more luminous
  than the average of other objects in the $H$ band, one would not
  consider them significantly overluminous.  The $H$-band point for
  SN~2009dc corresponds to the same epoch as in the $J$-band plot,
  but SN~2009dc brightened 0.26 mag in $H$ over the following month.
  The diamond-shape points
  correspond to objects that peak in the near-IR later than $B$
  maximum.  These objects are subluminous in all bands.  See
  \citet{Kri_etal09} for more details.}
\end{figure}

\begin{figure}
\plotone{spectra.eps}
{\center Krisciunas {\it et al.} Figure \ref{spectra}.
Temporal sequence of spectra of
  SN~2001ay. The labels indicate the number of observer-frame days
  with respect to $B$-band maximum.  Eleven which have no telescope
  indicated were taken with the FLWO 1.5-m.  The $t$ = +6~d spectrum
  from the KPNO 2.1-m telescope has been combined with the $t$ = +9~d
  {\it HST} spectrum. The solid vertical line marks the nominal
  wavelength of the $\lambda_0$ = 6355~\AA\ line of Si~II. The dashed
  vertical line shows Si~II line blueshifted by 10,000 \kms.  Telluric
  oxygen is identified by an Earth symbol.}
\end{figure}

\begin{figure}
\plotone{additional_spectra.eps}
{\center Krisciunas {\it et al.} Figure \ref{additional_spectra}.
Additional spectra of SN~2001ay
  obtained with the Lick Observatory 3-m Shane telescope, {\it HST},
  and the Las Campanas Observatory 2.5-m du Pont telescope.  The
  captioning within the figure and the meaning of the vertical lines
  are the same as in Figure \ref{spectra}. Telluric oxygen is
  identified by an Earth symbol.}
\end{figure}

\clearpage

\begin{figure}
\plotone{keck.eps}
{\center Krisciunas {\it et al.} Figure \ref{keck}.
A portion of our highest signal-to-noise ratio
  spectrum of SN~2001ay, taken with the Keck ESI on 2001 April 22.
  The dashed lines for Si~III and S~II correspond to blueshifts of 9000
  \kms.  The dashed magenta lines for C~II correspond to a blueshift
  of 12,000 \kms.  The dot-dashed red lines for Mg~II and Si~II
  correspond to a blueshift of 14,000 \kms.  The dotted line for Na~I
  is the rest wavelength.}
\end{figure}

\clearpage

\begin{figure}
\plotone{01ay_NaID.eps}
{\center Krisciunas {\it et al.} Figure \ref{sodium}.
Profile of the Na~I~D lines in our Keck
  ESI spectrum.  The Milky Way and host-galaxy components are clearly
  present.}
\end{figure}

\begin{figure}
\plotone{silicon.eps}
{\center Krisciunas {\it et al.} Figure \ref{silicon}.
Blueshifted velocity of two Si~II lines in
  the spectra of SN~2001ay around maximum light.  Within $1\sigma$ the
  gradient derived from the two lines is the same, about 200
  \kms\ d$^{-1}$.  This qualifies SN~2001ay to be a ``high velocity
  gradient'' object \citep{Ben_etal05}.}
\end{figure}

\begin{figure}
\plotone{bolom_lc.eps}
{\center Krisciunas {\it et al.} Figure \ref{bolom_lc}.
Bolometric luminosities ($L$) of SNe~2001ay,
  2007if \citep{Sca_etal10}, and 2009dc \citep{Tau_etal11},
  measured in erg s$^{-1}$.  For SN~2001ay we have scaled our
  integrated bolometric
  luminosity by a factor of 1.15 to account for flux not included in
  our photometric bandpasses.  We adopted a distance modulus of $m-M
  = 35.55$ mag from Table \ref{properties}.}
\end{figure}

\begin{figure}
\plotone{SN01ay.Edep.eps}
{\center Krisciunas {\it et al.} Figure \ref{edep}.
The best-fit radioactive decay energy deposition
function (middle solid line) to the UVOIR light curve (blue squares) of
SN~2001ay.  Here we adopt Arnett's Rule \citep{Arn82}, with $\alpha$ = 1.0,
which stipulates that the gamma-ray deposition matches the bolometric flux
at maximum light. The top solid line corresponds to the case of complete
trapping of $\gamma$ rays and positrons (i.e., $\tau \gg 1$), while the
bottom solid line is the case of complete $\gamma$-ray escape ($\tau \ll 1$).
The vertical dashed line indicates the epoch at which the fit begins.
The best fit to the latest few points suggests that the ejecta became
optically thin to $\gamma$ rays 57.2 $\pm$ 2.4~d past maximum brightness.}
\end{figure}

\clearpage

\begin{figure}
\plotone{01ay_05eq_09dc.eps}
{\center Krisciunas {\it et al.} Figure \ref{3spectra}.
Comparison of the optical spectra of
  SNe~2001ay, 2005eq, and 2009dc.}
\end{figure}

\begin{figure}
\plotone{synow1.eps}
{\center Krisciunas {\it et al.} Figure \ref{synow1}.
SYNOW modeling of the optical spectrum of
  SN~2001ay at 1~d before $B$ maximum. Upper curve = data; lower
  curve = SYNOW model.  The dashed lines indicate the addition of
  C~II to the model.}
\end{figure}

\begin{figure}
\plotone{synow2.eps}
{\center Krisciunas {\it et al.} Figure \ref{synow2}.
SYNOW modeling of the optical spectrum of
  SN~2001ay 6~d after $B$ maximum.  Upper curve = data; lower
  curve = SYNOW model.}
\end{figure}

\begin{figure}
\plotone{synapps.eps}
{\center Krisciunas {\it et al.} Figure \ref{synapps}.
SYNAPPS modeling of the optical spectrum of
  SN~2001ay 1~d before $B$ maximum.  The blue dashed line (model
  spectrum) can be shifted arbitrarily along the vertical axis to match
  the actual spectrum (shown in black).}
\end{figure}

\end{document}